\documentclass[pra,floatfix]{revtex4}
\usepackage{amsmath}
\usepackage{amssymb}
\usepackage{graphicx}
\usepackage{dcolumn}
\usepackage{bm}

\begin{document}

\title{Comparison of Theory and Experiment \\
for a One-Atom Laser in a Regime of Strong Coupling}
\date{\today }
\author{A.~D.~Boozer, A.~Boca, J.~R.~Buck, J.~McKeever, and H.~J.~Kimble}
\affiliation{Norman Bridge Laboratory of Physics 12-33, California Institute of
Technology, Pasadena, CA 91125}

\begin{abstract}
Our recent paper reports the experimental realization of a
one-atom laser in a regime of strong coupling \cite{mckeever03b}.
Here we provide the supporting theoretical analysis relevant to
the operating regime of our experiment. By way of a simplified
four-state model, we investigate the passage from the domain of
conventional laser theory into the regime of strong coupling for a
single intracavity atom pumped by coherent external fields. The
four-state model is also employed to exhibit the vacuum-Rabi
splitting and to calculate the optical spectrum. We next extend
this model to incorporate the relevant Zeeman hyperfine states as
well as a simple description of the pumping processes in the
presence of polarization gradients and atomic motion. This
extended model is employed to make quantitative comparisons with
the measurements of Ref. \cite{mckeever03b} for the intracavity
photon number versus pump strength and for the photon statistics
as expressed by the intensity correlation function $g^{(2)}(\tau
)$.
\end{abstract}

\maketitle

\section{Introduction}

Although a number of theoretical analyses related to a one-atom
laser have appeared in the literature \cite
{mu92,ginzel93,pellizzari94a,pellizzari94b,horak95,briegel96,meyer97a,
loeffler97,meyer97b,meyer98,jones99,chough00,fidio01,kilin02,rice03,salzburger03},
these prior treatments have not been specific to the parameter
range of our recent experiment as reported in Ref.
\cite{mckeever03b}. Because of this circumstance, we have carried
out theoretical investigations in support of our experimental
program, and present comparisons of these model calculations with
our measurements in this paper. In Section II we introduce a
simplified four-state model that captures the essential features
of the operation of our one-atom laser in a domain of strong
coupling but which avoids the complexity of the full Zeeman
substructure of the hyperfine levels in atomic Cesium. Sections
III and IV then present in turn semiclassical and quantum
solutions for this four-state model system. By way of a physically
motivated transformation for which the length of a Fabry-Perot
cavity is made progressively shorter, we utilize these results to
investigate the continuous passage from a domain in which
conventional laser theory is applicable into a regime of strong
coupling for which the full quantum theory is required. We thereby
gain some insight into the relationship of our system to prior
theoretical treatments related to the definition of the laser
threshold and to \textquotedblleft
thresholdless\textquotedblright\ lasing \cite
{demartini88,rice94,jin94,bjork94,protsenko99}. The four-state
model is further employed to calculate the intracavity photon
number versus pump detuning, thereby exhibiting the
\textquotedblleft vacuum-Rabi\textquotedblright\ splitting for the
atom-cavity system \cite {eberly83,agarwal84,thompson92} and to
compute the optical spectrum of the intracavity field.

In Section V we describe the procedures followed to obtain
solutions for an expanded model that incorporates the relevant
Zeeman substructure for the Cesium atom ($32$ atomic states), two
modes of the cavity with orthogonal polarizations, and a simple
model to account for the polarization gradients of the optical
fields. Comparisons of the results from quantum jumps simulations
based upon this expanded model with our measurements of the mean
intracavity photon number $\bar{n}$ versus normalized pump
intensity $x$ (Figure 3 of Ref. \cite{mckeever03b}) and with our
experimental determination of the intensity correlation function
$g^{(2)}(\tau )$ (Figure 4 of Ref. \cite{mckeever03b}) are given
in Sections V(a) and V(b), respectively.

Our intent here is not to belabor the comparison of our experiment
with prior work on micro-masers and lasers, for which extensive
reviews are available \cite
{berman94,meystre92,yamamoto-slusher93,chang-campillo96,vahala03}.
Instead, our principal goal is to establish quantitative
correspondence between our measurements and fundamental
theoretical models. Having thereby validated the suitability of
the theoretical treatments, we can then use these models to inform
further experimental investigations of the atom-cavity system.

\section{Four-state model}

\begin{figure}[t]
\includegraphics[width=8cm]{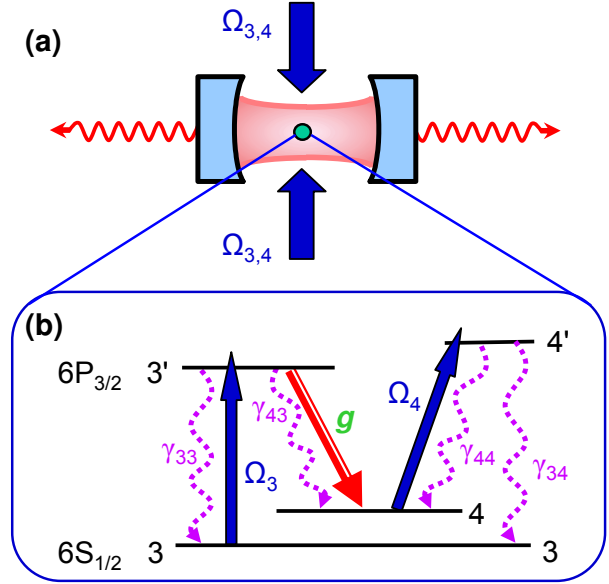}
\caption{Illustration of a one-atom laser. (a) The atom is located
in a high-$Q$ optical cavity of decay rate $\protect\kappa $, and
is driven by the fields $\Omega _{3,4}$. (b) Inset of the atomic
level scheme relevant to our experiment with the
$6S_{1/2}\leftrightarrow 6P_{3/2}$ transition in atomic Cesium.
The \textquotedblleft lasing\textquotedblright\ transition is from
the excited level $F=3^{\prime }$ to the ground level $F=4$.
Pumping of the excited $3^{\prime }$ level is by way of coherent
excitation from a laser with Rabi frequency $\Omega _{3}$.
Effective decay from the ground $4$ level is provided by the
combination of a second field with Rabi frequency $\Omega _{4}$
and spontaneous decay $4^{\prime }\rightarrow 3$. Various
radiative decay rates $\protect\gamma _{ij}$ appropriate to the
$D_{2}$ line in Cs are given in the text.} \label{setup}
\end{figure}

We begin with a four-state model to describe our experiment in
which a single Cesium atom is trapped inside an optical cavity as
illustrated in Figure \ref{setup}. Although the actual level
structure of the Cesium $ 6S_{1/2}\leftrightarrow 6P_{3/2}$
transition is more complex due to the Zeeman substructure, this
simpler model offers considerable insight into the nature of the
steady states and dynamics. Following the labelling convention in
Fig. \ref{setup}, we introduce the following set of Hamiltonians
$H_{i}$ in a suitably defined interaction picture ($\hslash =1$):
\begin{eqnarray}
\hat{H}_{1} &=&g_{43}(\hat{a}^{\dagger
}\hat{\sigma}_{g4,e3}+\hat{\sigma}
_{e3,g4}\hat{a}),  \label{hamiltonians} \\
\hat{H}_{2} &=&\frac{1}{2}\Omega
_{3}(\hat{\sigma}_{g3,e3}+\hat{\sigma}
_{e3,g3}),  \notag \\
\hat{H}_{3} &=&\frac{1}{2}\Omega
_{4}(\hat{\sigma}_{g3,e3}+\hat{\sigma}
_{e3,g3}),  \notag \\
\hat{H}_{4} &=&(\Delta _{AC}+\Delta _{4})\hat{a}^{\dagger }\hat{a},  \notag
\\
\hat{H}_{5} &=&\Delta _{3}\hat{\sigma}_{e3,e3}+\Delta
_{4}\hat{\sigma}
_{e4,e4},  \notag \\
\hat{H}_{tot}
&=&\hat{H}_{1}+\hat{H}_{2}+\hat{H}_{3}+\hat{H}_{4}+\hat{H}_{5}
\text{ .}  \notag
\end{eqnarray}
In a standard convention, the atomic operators are
$\hat{\sigma}_{i,j}\equiv |i\rangle \langle j|$ for states
$(i,j)$, with the association of the $F=3,4$ ground and the
$F^{\prime }=3^{\prime },4^{\prime }$ levels with $ g3,g4,e3,e4 $,
respectively. The Hamiltonian $\hat{H}_{1}$ accounts for the
coherent coupling of the atomic transition $e3\leftrightarrow g4$
to the field of a single mode of the cavity with creation and
annihilation operators $(\hat{a}^{\dagger },\hat{a})$. The upper
state $e3$ of the lasing transition is pumped by the
(coherent-state) field $\Omega _{3}$, while the lower state $g4 $
is depleted by the field $\Omega _{4}$ as described by $(
\hat{H}_{2},\hat{H}_{3})$, respectively.
$(\hat{H}_{4},\hat{H}_{5})$ account for various detunings,
including $\Delta _{AC}$ for the offset between the cavity
resonance and the $e3\leftrightarrow g4$ atomic transition,
$\Delta _{3}$ for the offset between the field $\Omega _{3}$ and
the $ g3\leftrightarrow e3$ transition, and $\Delta _{4}$ for the
offset between the field $\Omega _{4}$ and the $g4\leftrightarrow
e4$ transition. Beyond these interactions, we also account for
irreversible processes by assuming that the atom is coupled to a
continuum of modes other than the privileged cavity mode, and
likewise for the coupling of the cavity mode to an independent
continuum of external modes.

With these preliminaries, it is then straightforward to derive a master
equation for the density operator $\hat{\rho}$ for the atom-cavity system
\cite{carmichael-book,gardiner-book} in the Born-Markov approximation. For
our model system, this equation is
\begin{equation}
\frac{d\hat{\rho}}{dt}=-i[\hat{H}_{tot},\hat{\rho}]+\sum_{i=1}^{5}\hat{L}
_{i},  \label{master}
\end{equation}
Here, the terms $\hat{L}_{i}$ account for each of the various
decay channels, and are given explicitly by
\begin{eqnarray}
\hat{L}_{1} &=&\kappa (2\hat{a}\hat{\rho}\hat{a}^{\dagger }-\hat{a}^{\dagger
}\hat{a}\hat{\rho}-\hat{\rho}\hat{a}^{\dagger }\hat{a}),  \label{Li} \\
\hat{L}_{2} &=&\gamma
_{33}(2\hat{\sigma}_{g3,e3}\hat{\rho}\hat{\sigma}
_{e3,g3}-\hat{\sigma}_{e3,e3}\hat{\rho}-\hat{\rho}\hat{\sigma}_{e3,e3}),
\notag \\
\hat{L}_{3} &=&\gamma
_{43}(2\hat{\sigma}_{g4,e3}\hat{\rho}\hat{\sigma}
_{e3,g4}-\hat{\sigma}_{e3,e3}\hat{\rho}-\hat{\rho}\hat{\sigma}_{e3,e3}),
\notag \\
\hat{L}_{4} &=&\gamma
_{34}(2\hat{\sigma}_{g3,e4}\hat{\rho}\hat{\sigma}
_{e4,g3}-\hat{\sigma}_{e4,e4}\hat{\rho}-\hat{\rho}\hat{\sigma}_{e4,e4}),
\notag \\
\hat{L}_{5} &=&\gamma
_{44}(2\hat{\sigma}_{g4,e4}\hat{\rho}\hat{\sigma}
_{e4,g4}-\hat{\sigma}_{e4,e4}\hat{\rho}-\hat{\rho}\hat{\sigma}_{e4,e4})\text{
,}  \notag
\end{eqnarray}
where the association of each term $\hat{L}_{i}$\ with the decay processes
in Fig. \ref{setup} should be obvious. Spontaneous decay of the various
atomic transitions to modes other than the cavity mode proceeds at
(amplitude) rate $\gamma _{ij}$ as indicated in Fig. \ref{setup}, while the
cavity (field) decay rate is given by $\kappa $.

The master equation allows us to derive a set of equations for
expectation values of atom $\langle \hat{\sigma}_{i,j}\rangle $
and field $\langle \hat{a }\rangle $\ operators. One example is
for the atomic polarization $\langle \hat{\sigma}_{g4,e3}\rangle $
on the $e3\leftrightarrow g4$ transition, namely
\begin{eqnarray}
\frac{d\langle \hat{\sigma}_{g4,e3}\rangle }{dt} &=&-\left[ (\gamma
_{33}+\gamma _{43})+i\Delta _{3}\right] \langle \hat{\sigma}_{g4,e3}\rangle
\label{sg3e3} \\
&&-i\left( \Omega _{3}\langle \hat{\sigma}_{g4,g3}\rangle -\Omega
_{4}\langle \hat{\sigma}_{e4,e3}\rangle \right)  \notag \\
&&+ig_{43}\left( \langle \hat{\sigma}_{e3,e3}\hat{a}\rangle
-\langle \hat{ \sigma}_{g4,g4}\hat{a}\rangle \right) \text{ .}
\notag
\end{eqnarray}

A solution to this equation requires not only knowledge of
single-operator expectation values $\langle
\hat{\sigma}_{i,j}\rangle $ and $\langle \hat{a} \rangle $, but
also of operator products such as $\langle \hat{\sigma}
_{e3,e3}\hat{a}\rangle $. We can develop coupled equations for
such products $\langle \hat{\sigma}_{i,j}\hat{a}\rangle $ but
would find that their solution requires in turn yet higher order
correlations, ultimately leading to an unbounded set of equations.

Conventional theories of the laser proceed beyond this impasse by
one of several ultimately equivalent avenues. Within the setting
of our current approach, a standard way forward is to factorize
operator products in the fashion
\begin{equation}
\langle \hat{\sigma}_{i,j}\hat{a}\rangle =\langle \hat{\sigma}_{i,j}\rangle
\langle \hat{a}\rangle +(\langle \hat{\sigma}_{i,j}\hat{a}\rangle -\langle
\hat{\sigma}_{i,j}\rangle \langle \hat{a}\rangle )  \label{factor}
\end{equation}
with then the additional terms of the form $(\langle
\hat{\sigma}_{i,j}\hat{a }\rangle -\langle
\hat{\sigma}_{i,j}\rangle \langle \hat{a}\rangle )$ treated as
Langevin noise. Such approaches rely on system-size expansions in
terms of the small parameters $(1/n_{0},1/N_{0})$, where $
(n_{0},N_{0})$ are the critical photon and atom number introduced
in Ref. \cite{mckeever03b} for our one-atom laser. Within the
context of conventional laser theory, these parameters are
described more fully in Ref. \cite{carmichael-book,gardiner-book},
while their significance in cavity QED is discussed more
extensively in Ref. \cite{hjk sweden}. In qualitative terms,
conventional theories of the laser in regimes for which $
(n_{0},N_{0})\gg 1$ result in dynamics described by evolution of
mean values $\langle \hat{\sigma}_{i,j}\rangle $ and $\langle
\hat{a}\rangle $ (that are of order unity when suitably scaled),
with then small amounts of quantum noise (that arise from higher
order correlations of order $(1/n_{0},1/N_{0})\ll 1$ ).

In the following section, we discuss the so-called semiclassical solutions
obtained from the factorization $\langle \hat{\sigma}_{i,j}\hat{a}\rangle
=\langle \hat{\sigma}_{i,j}\rangle \langle \hat{a}\rangle $ neglecting
quantum noise. In Section IV, we then describe the full quantum solution
obtained directly from the master equation.

\section{Semiclassical theory for a four-state atom}

We will not present the full set of semiclassical equations here
since they are derived in a standard fashion from the master
equation Eq. \ref{master} \cite{sargent-book,gardiner-book}. One
example is for the atomic polarization $\langle
\hat{\sigma}_{g4,e3}\rangle $ on the $ e3\leftrightarrow g4$
transition, for which Eq. \ref{sg3e3} becomes
\begin{eqnarray}
\frac{d\langle \hat{\sigma}_{g4,e3}\rangle }{dt} &=&-\left[ (\gamma
_{33}+\gamma _{43})+i\Delta _{3}\right] \langle \hat{\sigma}_{g4,e3}\rangle
\label{sg3e3sc} \\
&&-i\left( \Omega _{3}\langle \hat{\sigma}_{g4,g3}\rangle -\Omega
_{4}\langle \hat{\sigma}_{e4,e3}\rangle \right)  \notag \\
&&+ig_{43}\left( \langle \hat{\sigma}_{e3,e3}\rangle -\langle
\hat{\sigma} _{g4,g4}\rangle \right) \alpha \text{ ,}  \notag
\end{eqnarray}
where $\alpha \equiv \langle \hat{a}\rangle $. There is a set of
$18$ such equations for the real and imaginary components of the
various field and atomic operators, together with the constraint
that the sum of populations over the four atomic states be unity.
We obtain the steady state solutions to these equations, where for
the present purposes, we restrict attention to the case of zero
detunings $\Delta _{AC}=\Delta _{3}=\Delta _{4}=0$. Allowing for
nonzero detunings of atom and cavity would add to the complexity
of the semiclassical analysis because of the requirement for the
self-consistent solution for the frequency of emission [see, for
example, Ref. \cite{eschmann99} for the case of a (multi-atom)
Raman laser].

The semiclassical solutions are obtained for the parameters relevant to our
experiment with atomic Cs, namely
\begin{equation}
(\gamma _{33},\gamma _{43},\gamma _{44},\gamma
_{34})=(\frac{3}{4},\frac{1}{4 },\frac{7}{12},\frac{5}{12})\gamma
\text{ ,}  \label{gamma}
\end{equation}
where these rates are appropriate to the (amplitude) decay of the
levels $ 6P_{3/2},F^{\prime }=3^{\prime },4^{\prime }\rightarrow
6S_{1/2},F=3,4$ with $\gamma =2\pi \times 2.6$ MHz (i.e., a
radiative lifetime $\tau =1/2\gamma =30.6$ ns). The cavity (field)
decay rate $\kappa $ is measured to be $ \kappa =2\pi \times 4.2$
MHz. The rate of coherent coupling $g_{43}$ for the
$e3\leftrightarrow g4$ transition (i.e., $6P_{3/2},F^{\prime
}=3^{\prime }\leftrightarrow 6S_{1/2},F=4$) is calculated from the
known cavity geometry (waist and length) and the decay rate
$\gamma $, and is found to be $ g_{43}=2\pi \times 16$ MHz based
upon the effective dipole moment of the transition.

\begin{figure*}[t]
\includegraphics[width=16cm]{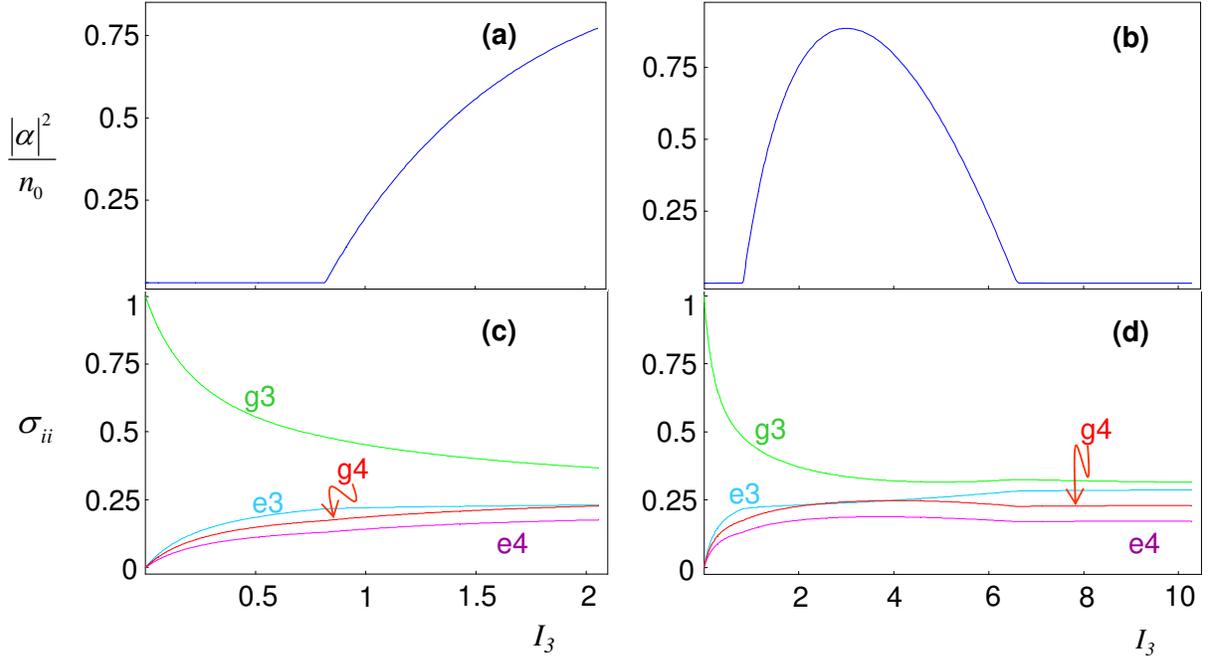}
\caption{Results from the semiclassical theory as applied to the
atom-cavity system in Fig. 1. (a,b) Intracavity intensity
$|\protect\alpha |^{2}$ in units of the critical photon number
$n_{0}$ is plotted as a function of the pump intensity
$I_{3}=(\frac{\Omega _{3}}{2 \protect\gamma })^{2}$. A threshold
for $|\protect\alpha |^{2}$ is evident for $I_{3}\simeq 0.8$.
(c,d) Populations $\protect\sigma _{ii}=\langle
\hat{\protect\sigma} _{ii}\rangle $ versus $I_{3}$. In (c),
population inversion $\protect\sigma _{e3,e3}>\protect\sigma
_{g4,g4}$ occurs over a wide range as the pump intensity $I_{3}$
is increased from $0$, including in the threshold region $
I_{3}\simeq 0.8$, with then \textquotedblleft population
clamping\textquotedblright\ for $\protect\sigma _{e3,e3}$ as
$I_{3}$ increases beyond threshold. In all cases, the recycling
intensity $I_{4}=( \frac{\Omega _{4}}{2 \protect\gamma })^{2}=3$
and the detunings $\Delta _{AC}=\Delta _{3}=\Delta _{4}=0$.}
\label{semiclassical}
\end{figure*}

Examples of the resulting steady-state solutions for the
intracavity intensity $|\alpha |^{2}$ together with the
populations $\sigma _{ii}$ of the four atomic states are displayed
in Figure \ref{semiclassical}. Parts (a) and (c) of the figure
illustrate the behavior of $|\alpha |^{2}$ and $ \sigma _{ii}$
around the semiclassical threshold as functions of the pump
intensity $I_{3}$. Parts (b) and (d) explore these dependencies
over a wider range in $I_{3}$. For fixed ratios among the various
decay rates as in Eq. \ref{gamma}, the semiclassical solutions for
$|\alpha |^{2}/n_{0}$ as well as the various populations $\sigma
_{ii}$ plotted in Fig. \ref{semiclassical} depend only on the
critical atom number $N_{0}$ (or equivalently, the cooperativity
parameter $C_{1}=1/N_{0}$ for a single atom in the cavity). Hence,
as emphasized in the \textit{Supplementary Information} published
with our paper Ref. \cite{mckeever03b}, these steady state
solutions from the semiclassical theory are independent of the
cavity length $l$, and provide a point of reference for
understanding \textquotedblleft lasing\textquotedblright\ for a
single atom in a cavity. This is because $ N_{0}=\frac{2\kappa
\gamma }{g^{2}}$ is independent of cavity length $l$ for a cavity
with constant mirror reflectivity and cavity waist $w_{0}$.

Importantly, the semiclassical theory predicts threshold behavior
for parameters relevant to our experiment, including inversion $
\sigma _{e3,e3}>\sigma _{g4,g4}$ in the threshold region, although
this is not essential for Raman gain for $g3\rightarrow g4$ via
$e3$. One atom in a cavity can exhibit such a \textquotedblleft
laser\textquotedblright\ transition for the steady state solutions
in the semiclassical theory because the cooperativity parameter
$C_{1}\gg 1$. Indeed, in these calculations we used our
experimental value for the cooperativity parameter $
C_{1}=1/N_{0}\simeq 12$. Among other relevant features illustrated
in Fig. \ref{semiclassical} is the quenching of the laser emission
around $ I_{3}\simeq 6.5$, presumably due to an Autler-Townes
splitting of the excited state $e3$ at high pump intensity
\cite{meyer97a}.

\subsection{Relationship to a Raman laser}

In many respects our system is quite similar to a three-level
Raman scheme, for which there is an extended literature (e.g.,
Ref. \cite{eschmann99} and references therein). In fact we have
carried out an extensive analysis of a Raman scheme analogous to
our system in Fig. \ref{setup}. Pumping is still done by the field
$\Omega _{3}$ on the $3\rightarrow 3^{\prime }$ transition.
However, recycling $4\rightarrow 4^{\prime }\rightarrow 3$ by the
field $\Omega _{4}$ and decay $\gamma _{34}$ is replaced by direct
decay $4\rightarrow 3$ at a fictitious incoherent rate of decay $
\beta _{34}$ with level $4^{\prime }$ absent. In all essential
details, the results from this analysis are in correspondence with
those presented from our four-level analysis in this section. In
particular, the threshold onsets in precisely the same fashion as
in Fig. \ref{semiclassical}(a), and the output is
\textquotedblleft extinguished\textquotedblright\ at high pump
levels for $\Omega _{3}$. This turn-off appears to be associated
with an AC-Stark splitting of the excited $3^{\prime }$ level by
the $\Omega _{3}$ field that drives the $3^{\prime }\rightarrow 4$
level out of resonance with the cavity due to the splitting of the
upper level $3^{\prime }$. Over the range of intensities explored
in this section, the \textquotedblleft
quenching\textquotedblright\ behavior seems to be unrelated to any
coherence effect associated with the combination of the field
$\Omega _{4}$ and decay $ \gamma _{34}$.

\section{Quantum theory for a four-state atom}

A one-atom laser operated in a regime of strong coupling has
characteristics that are profoundly altered from the familiar case
(described e.g. in Refs. \cite{carmichael-book,gardiner-book}),
for which the semiclassical equations are supplemented with
(small) quantum noise terms. The question then arises as how to
recognize a laser in this new regime of strong coupling, where we
recall the difficulty that this issue engenders even for systems
with critical photon number much greater than unity \cite
{rice94,jin94,bjork94,protsenko99}. The perspective that we adopt
here is to investigate the continuous transformation of a one-atom
laser from a domain of weak coupling for which the conventional
theory should be approximately valid into a regime of strong
coupling for which the full quantum theory is required.

\begin{figure}[b]
\includegraphics[width=8cm]{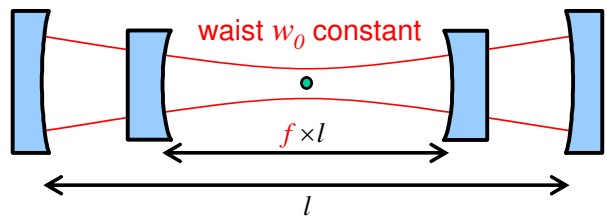}
\caption{Illustration of the scaling transformation considered in
Eqn. \protect\ref{scale} whereby the length of a spherical mirror
Fabry-Perot cavity is transformed $l\rightarrow fl$ while the
cavity waist $w_{0}$ and the atomic position are held constant.
The atom is indicated by the \textquotedblleft
dot\textquotedblright\ in the center of the cavity mode.}
\label{scaling}
\end{figure}

Towards this end, we consider a scenario in which the cavity
length (and hence its volume) is gradually reduced from a
\textquotedblleft large\textquotedblright\ value for which the
conventional theory is valid to a \textquotedblleft
small\textquotedblright\ value for which the system is well into a
regime of strong coupling. As illustrated in Figure \ref{scaling}
, this transformation is assumed to be under conditions of
constant cavity waist $w_{0}$ and mirror reflectivity $R$, in
which case scaling the length by a factor $f$ causes the other
parameters to scale as follows:
\begin{eqnarray}
l &\rightarrow &l_{f}=fl,  \label{scale} \\
g &\rightarrow &g_{f}=g/f^{1/2},  \notag \\
\kappa &\rightarrow &\kappa _{f}=\kappa /f,  \notag \\
\gamma &\rightarrow &\gamma ,  \notag \\
N_{0} &\rightarrow &N_{0},  \notag \\
n_{0} &\rightarrow &n_{0f}=fn_{0}\text{ .}  \notag
\end{eqnarray}
Recall that in the semiclassical theory illustrated in Fig. \ref
{semiclassical}, the quantity $|\alpha |^{2}/n_{0f}$ is invariant
under this transformation. By contrast, the role of single photons
becomes increasingly important as the cavity length is reduced
(i.e., $ n_{0f} $ becomes ever smaller), so that deviations from
the familiar semiclassical characteristics should become more
important, and eventually dominant.

\begin{figure*}[htb]
\includegraphics[width=16cm]{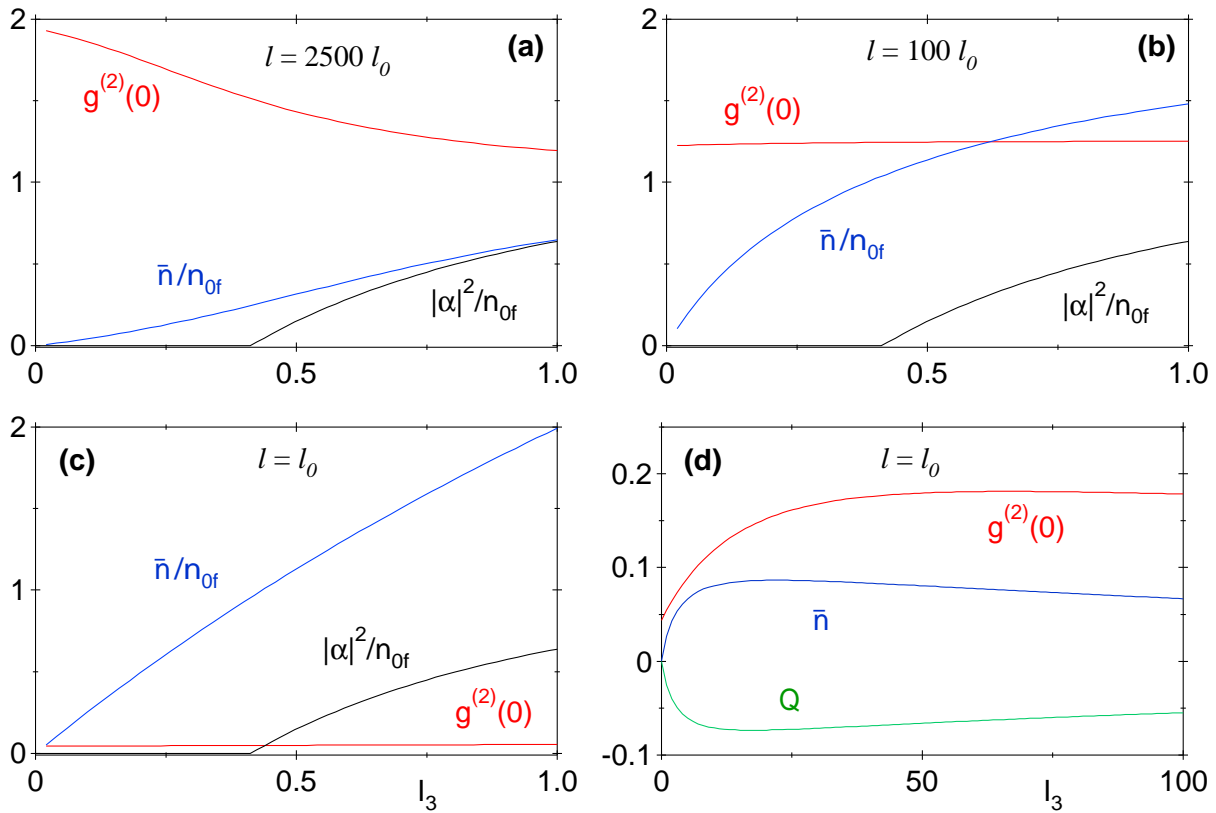}
\caption{The mean intracavity photon number $\bar{n}/n_{0f}$
(blue) and normalized intensity correlation function $g^{(2)}(0)$
(red) are plotted as functions of pump intensity $I_{3}=(\Omega
_{3}/2\protect\gamma )^{2}$ in (a)-(d). In (a)-(c), the cavity
length is made progressively shorter ($ 2500l_{0}$, $100l_{0}$,
$l_{0}$), where $l_{0}=42.2~\mathrm{\protect\mu m}$ is the length
of our actual cavity. The corresponding saturation photon numbers
are $n_{0f}=$($33.0$, $1.32$, $0.013$). $\bar{n}/n_{0f}$ and $
g^{(2)}(0)$ are calculated from the quantum theory for the
four-state system in Fig. \protect\ref{setup}, while
$|\protect\alpha |^{2}/n_{0f}$ given by the black curve is from
the semiclassical theory. (d) $\bar{n}$ (blue), $ g^{(2)}(0)$
(red), and the Mandel $Q$ parameter (green) shown over an extended
range of pump intensity $I_{3}$ for $l=l_{0}$. In all cases, $
I_{4}=(\Omega _{4}/2\protect\gamma )^{2}=2$, the $3\rightarrow
4^{\prime }$ and $4\rightarrow 4^{\prime }$ transitions are driven
on resonance, and the cavity detuning $\protect\omega _{CA}=0$.
Other parameters are as given in the text.} \label{n-g2vsi3}
\end{figure*}

\subsection{Field and atom variables for various cavity lengths}

\begin{figure*}[htb]
\includegraphics[width=16cm]{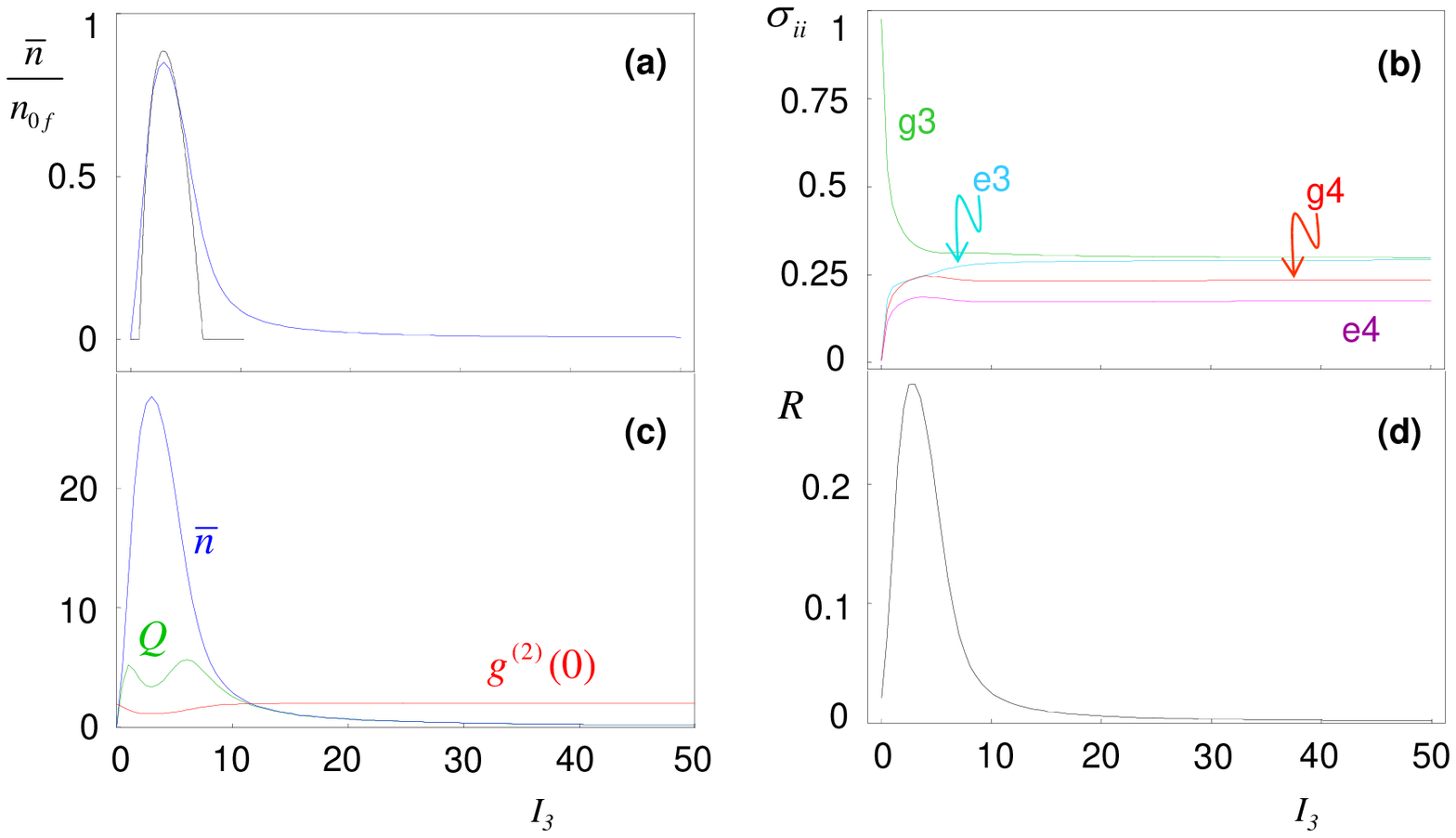}
\caption{Steady state solutions as functions of pump intensity
$I_{3}$ obtained from the numerical solution of the master
equation \protect\ref{master} for the four-state atom in a cavity
illustrated in Fig. \protect\ref{setup}. Here, the cavity length
$l=2500l_{0}$, where $l_{0}=42.2\protect\mu $m is the cavity
length in our experiment. (a) Mean intracavity photon number $
\bar{n}$ normalized to the saturation photon number $n_{0f}=33$
(in blue). The corresponding result for $|\protect\alpha
|^{2}/n_{0f}$ from the semiclassical theory is given by the black
curve. (b) Populations $\protect \sigma _{ii}$ of the four states
as labelled. (c) Mean intracavity photon number $\bar{n}$ (blue),
Mandel $Q$ parameter (green), and intensity correlation function
$g^{(2)}(0)$ (red). (d) Ratio $R$\ of photon flux from the cavity
mode $\protect\kappa _{f}$ $\bar{n}$ as compared to the rate of
atomic fluorescence $\protect\gamma _{43}\protect\sigma _{e3,e3}$
for the excited state $e3$.\ In all cases, the depleting intensity
$I_{4}=(\frac{ \Omega _{4}}{2\protect\gamma })^{2}=3$ and the
detunings $\Delta _{AC}=\Delta _{3}=\Delta _{4}=0$. Field and atom
decay rates are as specified in the text.} \label{f_0p02}
\end{figure*}

\begin{figure*}[htb]
\includegraphics[width=16cm]{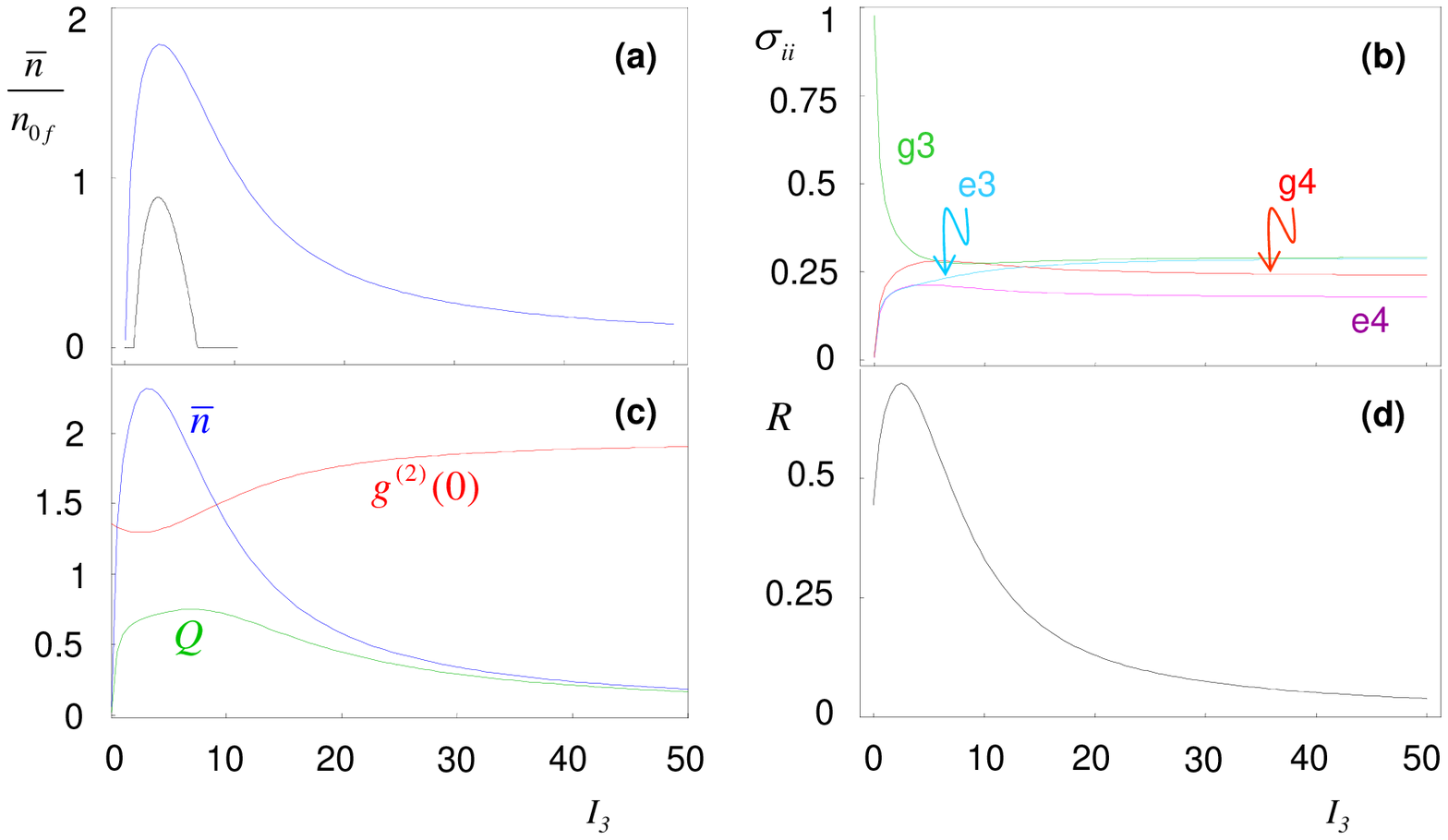}
\caption{Steady state solutions as functions of pump intensity
$I_{3}$ obtained from the numerical solution of the master
equation \protect\ref{master} for the four-state atom in a cavity
illustrated in Fig. \protect\ref{setup}. Here, the cavity length
$l=100l_{0}$, where $l_{0}=42.2\protect\mu $m is the cavity length
in our experiment. (a) Mean intracavity photon number $\bar{n}$
normalized to the saturation photon number $n_{0f}=1.3$ (in blue).
The corresponding result for $|\protect\alpha |^{2}/n_{0f}$ from
the semiclassical theory is given by the black curve. (b)
Populations $\protect \sigma _{ii}$ of the four states as
labelled. (c) Mean intracavity photon number $\bar{n}$ (blue),
Mandel $Q$ parameter (green), and intensity correlation function
$g^{(2)}(0)$ (red). (d) Ratio $R$\ of photon flux from the cavity
mode $\protect\kappa _{f}$ $\bar{n}$ as compared to the rate of
atomic fluorescence $\protect\gamma _{43}\protect\sigma _{e3,e3}$
for the excited state $e3$.\ In all cases, the depleting intensity
$I_{4}=(\frac{ \Omega _{4}}{2\protect\gamma })^{2}=3$ and the
detunings $\Delta _{AC}=\Delta _{3}=\Delta _{4}=0$. Field and atom
decay rates are as specified in the text.} \label{f_0p1}
\end{figure*}

\begin{figure*}[htb]
\includegraphics[width=16cm]{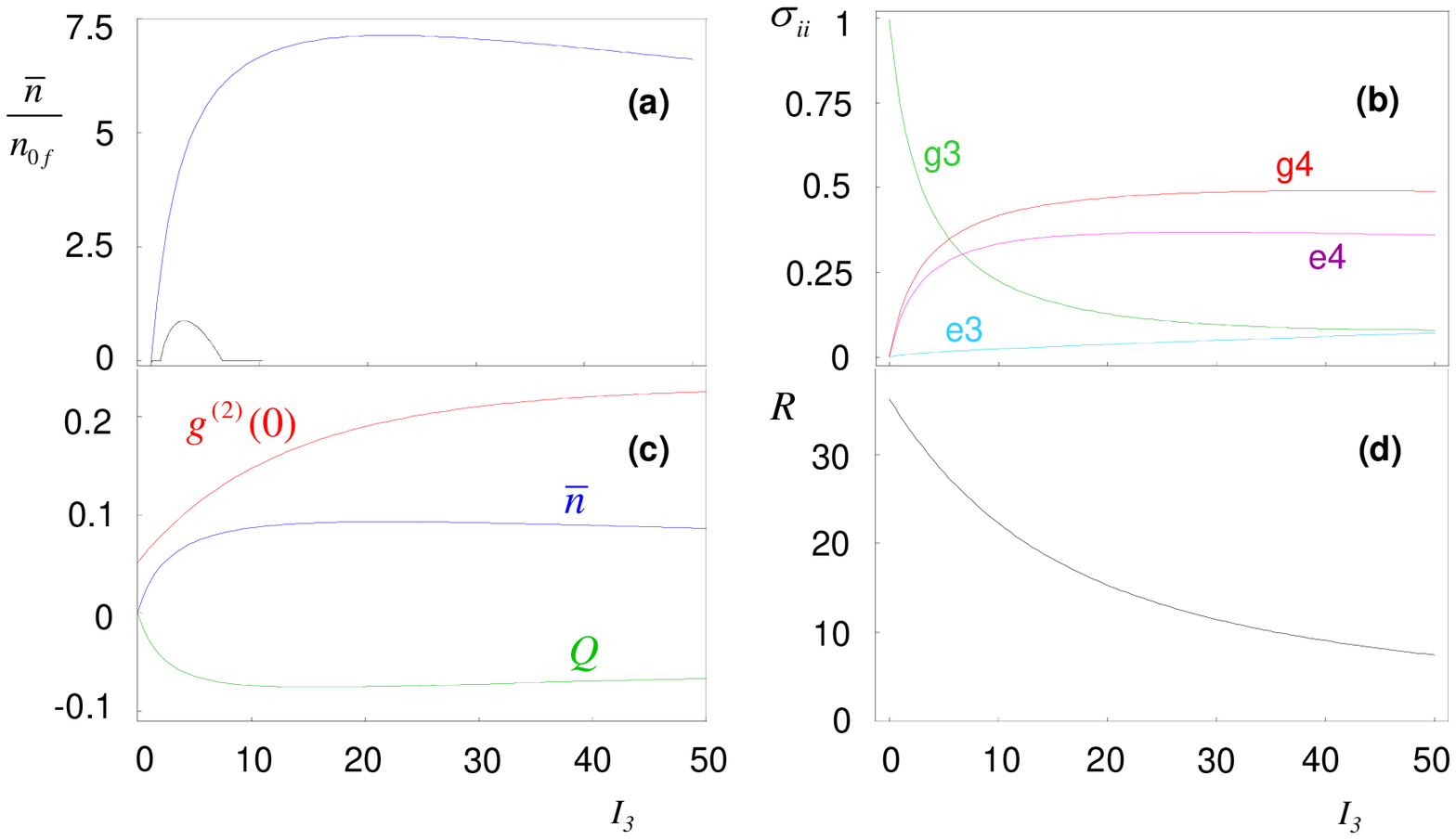}
\caption{Steady state solutions as functions of pump intensity
$I_{3}$ obtained from the numerical solution of the master
equation \protect\ref{master} for the four-state atom in a cavity
illustrated in Fig. \protect\ref{setup}. Here, the cavity length
$l=l_{0}$, where $l_{0}=42.2\protect\mu $m is the cavity length in
our experiment. (a) Mean intracavity photon number $\bar{n}$
normalized to the saturation photon number $n_{0}=0.013$ (in
blue). The corresponding result for $|\protect\alpha |^{2}/n_{0}$
from the semiclassical theory is given by the black curve. (b)
Populations $\protect \sigma _{ii}$ of the four states as
labelled. (c) Mean intracavity photon number $\bar{n}$ (blue),
Mandel $Q$ parameter (green), and intensity correlation function
$g^{(2)}(0)$ (red). (d) Ratio $R$\ of photon flux from the cavity
mode $\protect\kappa $ $\bar{n}$ as compared to the rate of atomic
fluorescence $\protect\gamma _{43}\protect\sigma _{e3,e3}$ for the
excited state $e3$.\ In all cases, the depleting intensity
$I_{4}=(\frac{ \Omega _{4}}{2\protect\gamma })^{2}=3$ and the
detunings $\Delta _{AC}=\Delta _{3}=\Delta _{4}=0$. Field and atom
decay rates are as specified in the text.} \label{f_1}
\end{figure*}

\begin{figure*}[htb]
\includegraphics[width=16cm]{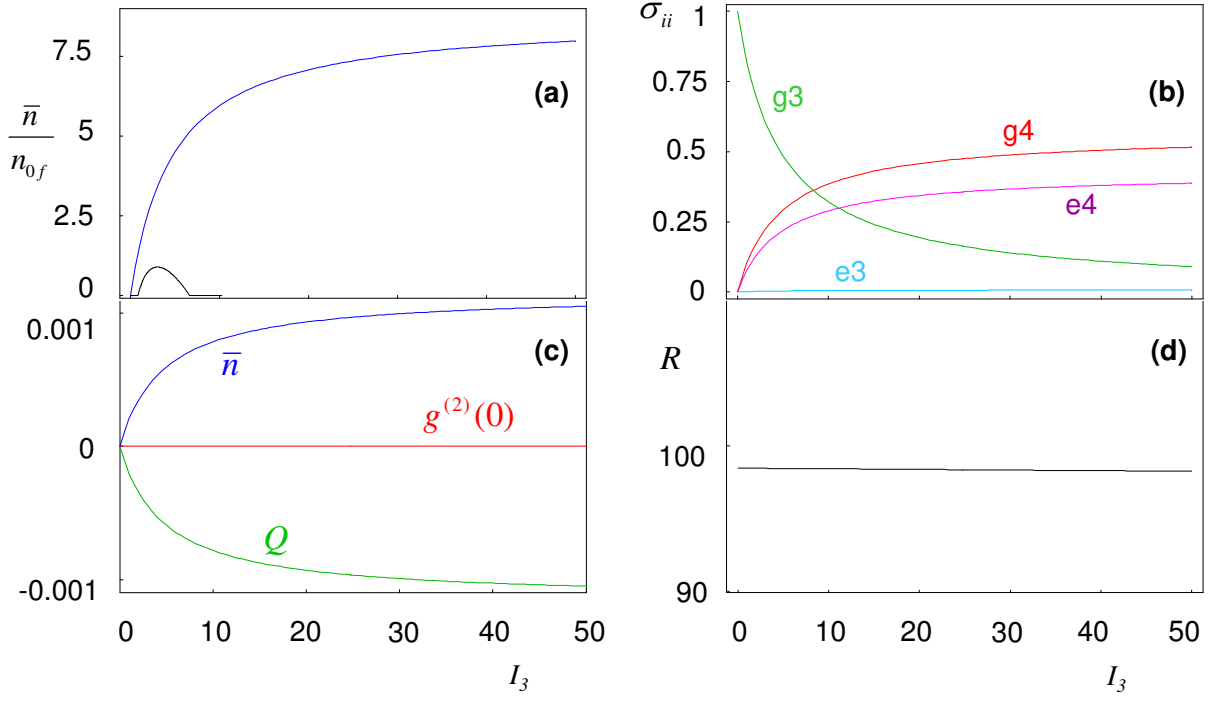}
\caption{Steady state solutions as functions of pump intensity
$I_{3}$ obtained from the numerical solution of the master
equation \protect\ref{master} for the four-state atom in a cavity
illustrated in Fig. \protect\ref{setup}. Here, the cavity length
$l=l_{0}/99\simeq \protect\lambda _{0}/2$ (i.e., $ f=1/99$), where
$l_{0}=42.2\protect\mu $m is the cavity length in our experiment
and $\protect\lambda _{0}=852.3$ nm is the wavelength of the
cavity QED transition. (a) Mean intracavity photon number
$\bar{n}$ normalized to the saturation photon number
$n_{0f}=1.31\times 10^{-4}$ (in blue). The corresponding result
for $|\protect\alpha |^{2}/n_{0f}$ from the semiclassical theory
is given by the black curve. (b) Populations $\protect \sigma
_{ii}$ of the four states as labelled. (c) Mean intracavity photon
number $\bar{n}$ (blue), Mandel $Q$ parameter (green), and
intensity correlation function $g^{(2)}(0)$ (red). (d) Ratio $R$\
of photon flux from the cavity mode $\protect\kappa _{f}$
$\bar{n}$ as compared to the rate of atomic fluorescence
$\protect\gamma _{43}\protect\sigma _{e3,e3}$ for the excited
state $e3$.\ In all cases, the depleting intensity $I_{4}=(\frac{
\Omega _{4}}{2\protect\gamma })^{2}=3$ and the detunings $\Delta
_{AC}=\Delta _{3}=\Delta _{4}=0$. Field and atom decay rates are
as specified in the text.} \label{f_sqrt99}
\end{figure*}

\begin{figure*}[htb]
\includegraphics[width=16cm]{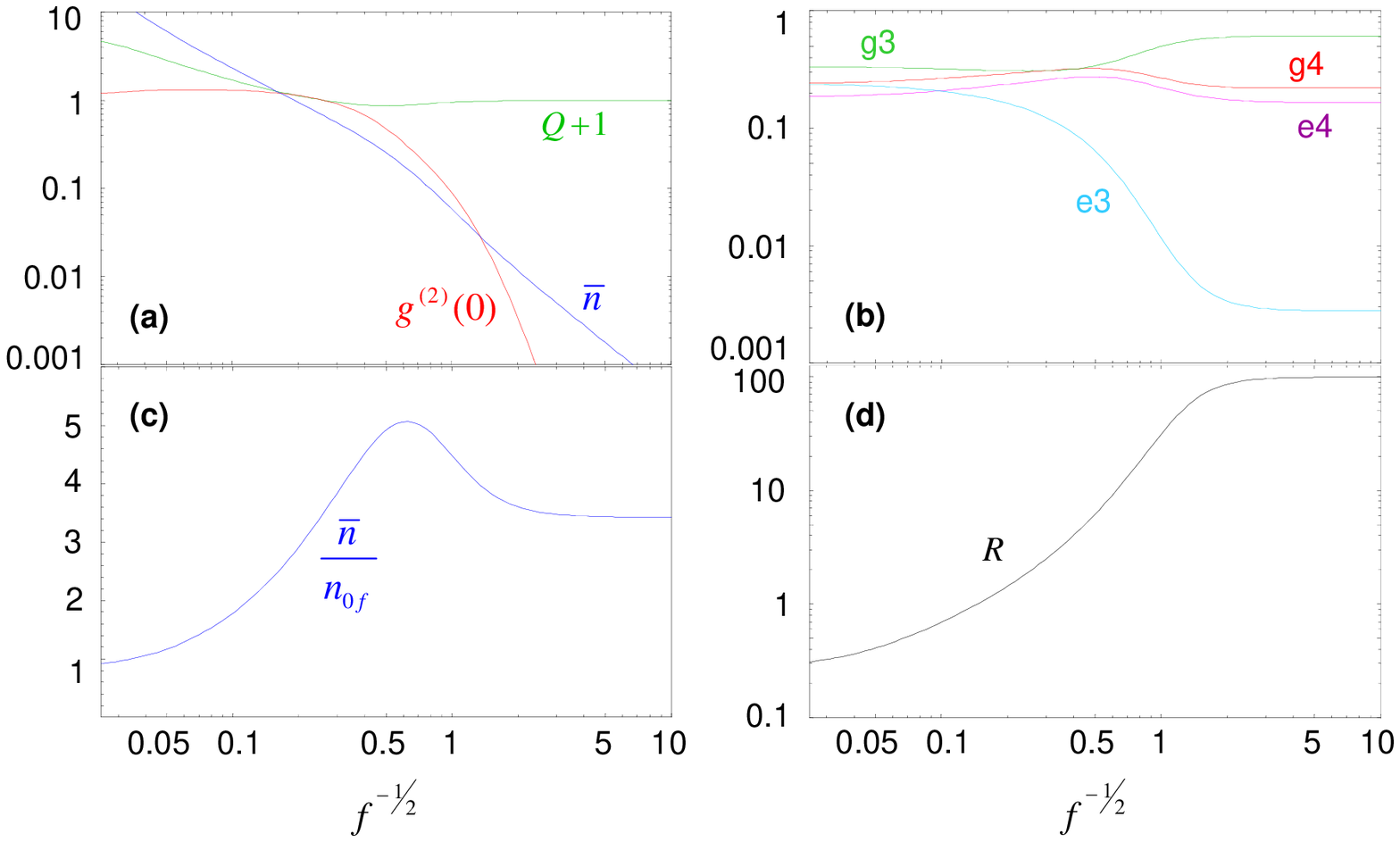}
\caption{Scaling behavior of various quantities as the cavity
length $l$ is varied, where $f=l/l_{0}$, and $l_{0}=42.2$
$\protect\mu $m for our actual cavity. Note that $g\varpropto
f^{-1/2}$ and $\protect\kappa \varpropto f^{-1}$ and that the
range in $f$ corresponds to that spanned by Figs.
\protect\ref{f_0p02} to \protect\ref{f_sqrt99}, namely
$0.01\lesssim f\lesssim 2500$. (a) Mean intracavity photon number
$\bar{n}$ (blue), the Mandel $Q$ parameter ($Q+1$ in green), and
the intensity correlation function $g^{(2)}(0)$ (red). (b)
Populations $\protect\sigma _{ii}$ of the four states as labelled.
(c) Mean intracavity photon number $\bar{n}$ normalized to the
saturation photon number $n_{0f}=n_{0}f=0.013\times f$. (d) Ratio
$R$\ of photon flux from the cavity mode $\protect\kappa
_{f}\bar{n }$ as compared to the rate of atomic fluorescence
$\protect\gamma _{43} \protect\sigma _{e3,e3}$ for the excited
state $e3$, where $\protect\kappa _{f}=\protect\kappa /f$. In all
cases, the pumping and recycling intensities $I_{3,4}=3$ and the
detunings $\Delta _{AC}=\Delta _{3}=\Delta _{4}=0$. Field and atom
decay rates are as specified in the text.} \label{variousvsf}
\end{figure*}

Framed by this perspective, we now present results from the
quantum treatment for a four-state model for the atom. Our
approach is to obtain steady state results for various operator
expectation values directly from numerical solutions of the master
equation given in Eq. \ref{master}\ by way of the \textit{Quantum
Optics Toolbox} written by S. Tan \cite{tan99}. Since such
numerical methods are by now familiar tools, we turn directly to
results from this investigation presented in Figs.
\ref{n-g2vsi3}-\ref{variousvsf}.

These figures display the behavior of various characteristics of
the atom-cavity system as the cavity length is reduced from
$l=2500l_{0}$ to $ l=100l_{0}$ to $l=l_{0}$ to $l=l_{0}/99$, where
$l_{0}=42.2\mu $m is the actual length of our cavity. Figure
\ref{n-g2vsi3} provides an overview of the evolution and is
reproduced from the \textit{Supplementary Information} in Ref.
\cite{mckeever03b}, while Figures \ref{f_0p02}-\ref{variousvsf}
provide more detailed information about the intracavity field and
atomic populations.

Figure \ref{n-g2vsi3}(a-c) and part (a) in Figs. \ref{f_0p02},
\ref{f_0p1}, and \ref{f_1} display the mean intracavity photon
number $\bar{n}/n_{0f}$ (where $n_{0f}$ is calculated for the
particular length), and compare this result to $|\alpha
|^{2}/n_{0f}$ from the semiclassical theory. The correspondence is
close in Figs. \ref{n-g2vsi3}(a) and \ref{f_0p02}(a) since
$n_{0f}=33$ in this case, but becomes increasingly divergent in
Figs. \ref {n-g2vsi3}(b) and \ref{f_0p1}(a) for which
$n_{0f}=1.3$, and in Figs. \ref {n-g2vsi3}(c) and \ref{f_1}(a) for
which $n_{0f}(f=1)=n_{0}=0.013$ (as in our experiment).

In qualitative terms, the peak in each of the curves for $\bar{n}/n_{0f}$ in
Figs. \ref{f_0p02}, \ref{f_0p1}, and \ref{f_1} arises because of a
\textquotedblleft bottleneck\textquotedblright\ in the cycle $g3\rightarrow
e3\rightarrow g4\rightarrow e4\rightarrow g3$. For our scheme with one atom
in a cavity, this cycle can proceed at a rate no faster than that set by the
decay rate $\gamma _{34}$. For higher pump intensities $I_{3}$, the
quenching of the emission displayed by the semiclassical theory becomes less
and less evident with decreasing $l$ as the coherent coupling rate $g$
becomes larger in a regime of strong coupling.

Part (b) in Figs. \ref{f_0p02}, \ref{f_0p1}, and \ref{f_1} shows the
populations $\sigma _{ii}$ of the four states. A noteworthy trend here is
the rapid reduction of the population $\sigma _{e3,e3}$ with decreasing
cavity length. Again, the rate $g$ becomes larger as $l$ is reduced, and
eventually overwhelms all other rates, so that population promoted to this
state is suppressed.

Figure \ref{n-g2vsi3} and part (c) in Figs. \ref{f_0p02},
\ref{f_0p1}, and \ref{f_1} address the question of the photon
statistics by plotting the Mandel $Q$ parameter (or equivalently
the Fano factor $F=Q+1$) as well as the normalized second-order
intensity correlation function $g^{(2)}(0)$ \cite
{mandel-wolf-95}. As shown in Fig. \ref{n-g2vsi3}(a), for large $
l=2500l_{0} $, the region around the semiclassical threshold
displays the familiar behavior associated with a conventional
laser \cite
{gardiner-book,mandel-wolf-95,sargent-book,haken-book,scully-zubairy},
namely that $g^{(2)}(0)$ evolves smoothly from $g^{(2)}(0)\approx
2$ below the semiclassical threshold to $g^{(2)}(0)\approx 1$
above this threshold. Furthermore, Fig. \ref{f_0p02}(c) shows that
the Mandel $Q$ parameter has a maximum in the region of the
threshold \cite{rice94}. Beyond this conventional (first)
threshold, the Mandel $Q$ parameter in Fig. \ref{f_0p02} (c) also
exhibits a second maximum, that has been described as a
\textquotedblleft second\textquotedblright\ threshold for one-atom
lasers \cite{meyer97a}, and $g^{(2)}(0)$ rises back from $1$ to
$2$. With decreasing cavity length, these features are lost as we
move into a regime of strong coupling. For example, the two peaks
in $Q$ merge into one broad minimum with $Q<0$ indicating the
onset of manifestly quantum or nonclassical character for the
emission from the atom-cavity system.

Finally, part (d) in Figs. \ref{f_0p02}, \ref{f_0p1}, and
\ref{f_1} presents results for the ratio $R$, where
\begin{equation}
R\equiv \frac{\kappa _{f}\bar{n}}{\gamma _{43}\sigma _{e3,e3}}
\label{R}
\end{equation}
gives the ratio of photon flux $\kappa _{f}\bar{n}$ \ from the
cavity mode to the photon flux $\gamma _{43}\sigma _{e3,e3}$
appearing as fluorescence into modes other than the cavity mode
from the spontaneous decay $e3\rightarrow g4$. For a conventional
laser, $\kappa _{f}\bar{n}\ll \gamma _{43}\sigma _{e3,e3}$ below
threshold, and $\kappa _{f}\bar{n}\gg \gamma _{43}\sigma _{e3,e3}$
above threshold, with the laser threshold serving as the abrupt
transition between these cases in the manner of a nonequilibrium
phase transition \cite{sargent-book,scully-zubairy}. As
illustrated in Fig. \ref{f_1}, no such transition is required in
the regime of strong coupling; $R\gg 1$ from the onset as the pump
$I_{3}$ is increased. This behavior is analogous to the
\textquotedblleft thresholdless\textquotedblright\ lasing
discussed in Refs. \cite {demartini88,jin94,bjork94,protsenko99}
and reviewed by Rice and Carmichael \cite{rice94}.

For the system illustrated in Figure \ref{scaling}, the
progression in length reduction has a limit at $l=\lambda _{0}/2$
corresponding to a Fabry-Perot cavity with length equal to the
lowest order longitudinal mode $ \lambda _{0}/2$, where $\lambda
_{0}=852.3$ nm is the wavelength of the cavity QED transition. To
reach this limit from the length $l_{0}$ appropriate to our actual
cavity, we must scale $l_{0}\rightarrow fl_{0}$ with $f=1/99$. In
a continuation of the sequence shown in Figs. \ref{f_0p02} ,
\ref{f_0p1}, and \ref{f_1}, we display in Fig. \ref{f_sqrt99}
results for such a cavity with $l=\lambda _{0}/2$. Note that
although $ C_{1}=1/N_{0}\simeq 12$ is invariant with respect to
this scaling and the saturation photon number is reduced to
$n_{0f}=1.31\times 10^{-4}$, nevertheless the atom-cavity system
has passed out of the domain of strong coupling, even though
$(n_{0f},N_{0})\ll 1$. This is because strong coupling requires
that $g_{0}\gg (\gamma ,\kappa )$, so that $(n_{0},N_{0})\ll 1$ is
a necessary but not sufficient condition for achieving strong
coupling. For the progression that we are considering with
diminishing length (but otherwise with the parameters of our
system), $l=\lambda _{0}/2$ does not lie within the regime of
strong coupling ($g_{43}/\gamma =61,g_{43}/\kappa =0.40$), but
rather more toward the domain of a \textquotedblleft
one-dimensional atom\textquotedblright , for which $\kappa \gg
g^{2}/\kappa \gg \gamma $ (see, for example, Refs. \cite
{lugiato,turchette95} for theoretical discussions and a previous
experimental investigation). In this domain of the Purcell effect
\cite {berman94,yamamoto-slusher93,chang-campillo96,vahala03}, the
fractional emission into the cavity mode as compared to
fluorescent emission into free space for the $3^{\prime
}\rightarrow 4$ transition is characterized by the parameter
\begin{equation}
\beta _{43}\equiv \frac{2C_{1}^{(43)}}{1+2C_{1}^{(43)}}\simeq 0.99\text{,}
\label{beta}
\end{equation}
where $C_{1}^{(43)}=C_{1}\times (\gamma /\gamma _{43})\simeq 48$.

As compared to Figs. \ref{f_0p02}, \ref{f_0p1}, and \ref{f_1}, a
noteworthy feature of the regime depicted in Fig. \ref{f_sqrt99}
is the absence of a dependence of $g^{(2)}(0)$ on the pump level
$I_{3}$. In fact, $ g^{(2)}(0)\simeq 0$ over the entire range
shown, so that the cavity field is effectively occupied only by
photon numbers $0$ and $1$. In correspondence to this situation,
the Mandel $Q$ parameter in Fig. \ref{f_sqrt99}(c) is essentially
given by the mean of the intracavity photon number, $Q\simeq -
\bar{n}$, with $\bar{n}\ll 1$. Furthermore, the dominance of
emission into the cavity mode over fluorescence decay becomes even
more pronounced than in Fig. \ref{f_1}(d), as documented by the
ratio $R$ in Fig. \ref{f_sqrt99}(d). In agreement with expectation
set by Eq. \ref{beta}, note that $R\simeq \beta _{43}/(1-\beta
_{43})$. All in all, the \textquotedblleft
bad-cavity\textquotedblright\ limit specified by $\kappa \gg
g^{2}/\kappa \gg \gamma $ \cite{turchette95,lugiato} (toward which
Fig. \ref{f_sqrt99} is pressing) is a domain of single-photon
generation for the atom-cavity system, which for $f\ll 1$ has
passed out of the regime of strong coupling.

Figures \ref{f_0p02}, \ref{f_0p1}, \ref{f_1}, and \ref{f_sqrt99}
provide a step-by-step description of the evolution of the
atom-cavity system from the domain of conventional laser theory
($l\gg l_{0}$ as in Fig. \ref{f_0p02} with $f=2500$), into the
regime of strong coupling ($l=l_{0}$ as in Fig. \ref {f_1} with
$f=1$), and then out of the strong-coupling regime into the
Purcell domain ($l=l_{0}/99\simeq \lambda _{0}/2$ as approached in
Fig. \ref {f_sqrt99} with $f=0.01$) \cite
{berman94,yamamoto-slusher93,chang-campillo96,vahala03}. We now
attempt to give a more global perspective of the scaling behavior
of the atom-cavity system by examining various field and atomic
variables directly as functions of the scale parameter
$f=l/l_{0}$. A particular set of such results is displayed in
Figure \ref{variousvsf}, where the pump intensity $I_{3}=3$ is
fixed near the peak in the output from the semiclassical theory in
Fig. \ref {semiclassical}, and the recycling intensity $I_{4}$ is
held constant at $ I_{4}=3$.

In Fig. \ref{variousvsf}(a) the mean intracavity photon number
$\bar{n}$ is seen to undergo a precipitous drop as the cavity
length is made progressively shorter (i.e., increasing $f^{-1/2}$,
since $l\varpropto f$). However, when $\bar{n}$ is normalized to
the critical photon number $n_{0f}$ , the quantity
$\bar{n}/n_{0f}$ is seen to approach unity for small $ f^{-1/2} $
(i.e., long cavities with $l\gg l_{0}$) as appropriate to the
conventional theory in Fig. \ref{f_0p02}). With increases in
$f^{-1/2}$ (i.e., shorter cavity lengths), $\bar{n}/n_{0f}$ rises
to a maximum around $ f\sim 3$ for strong coupling with $l\sim
l_{0}$ as in Fig. \ref{f_1}, before then decreasing to approach a
constant value for yet larger values of $ f^{-1/2}$ as the system
exits from the domain of strong coupling.

Also shown in Fig. \ref{variousvsf}(a) are the quantities
$g^{(2)}(0)$ and $ Q+1$ that characterize the photon statistics of
the intracavity field. As previously noted, $g^{(2)}(0)$ lies in
the range $1\leq g^{(2)}(0)\leq 2$ for conventional laser theory,
but drops below unity in the regime of strong coupling and
approaches zero for $f\ll 1$. In this same limit of very small
cavities in the Purcell regime, $Q\simeq -\bar{n}$.

Fig. \ref{variousvsf}(b) displays the populations for the
four-state system as functions of $f^{-1/2}$. For the conventional
regime with $f^{-1/2}\ll 1$ , there is population inversion,
$\sigma _{e3,e3}>\sigma _{g4,g4}$ (which was shown in Fig.
\ref{semiclassical} for small values of $I_{3}$), but this
possibility is lost for increasing $f^{-1/2}$ (i.e., decreasing
cavity length). Strong coupling dictates that the rate $g$
dominates all others, so that appreciable population cannot be
maintained in the state $e3$. Finally, Fig. \ref{variousvsf}(d)
displays the dependence of the ratio $R=(\kappa _{f}
\bar{n})/(\gamma _{43}\sigma _{e3,e3})$ on $f^{-1/2}$. From values
$R<1$ in the conventional domain, $R$ rises monotonically with
decreasing cavity length reaching the plateau $R\gg 1$\ specified
by Eq. \ref{beta}.

\subsection{Vacuum-Rabi splitting}

In the preceding discussion, we have compared various aspects of our
one-atom system with conventional lasers and have restricted the analysis to
the case of resonant excitation with $\Delta _{3}=0$. Our actual system
operates in a regime of strong coupling, so that there should be an explicit
manifestation of the \textquotedblleft vacuum-Rabi\textquotedblright\
splitting associated with one quantum of excitation in the $4\leftrightarrow
3^{\prime }$ manifold \cite{eberly83,agarwal84,thompson92}.

\begin{figure}[htb]
\includegraphics[width=8cm]{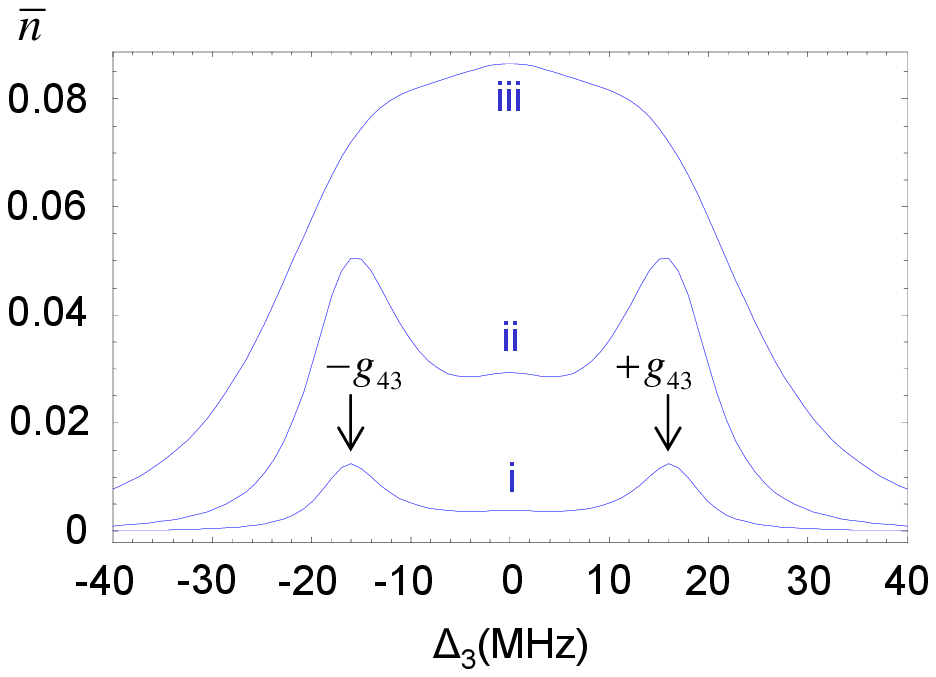}
\caption{The mean intracavity photon number $\bar{n}$ versus the
detuning $ \Delta _{3}$ (in cycles/sec) of the pump field $\Omega
_{3}$, where $\Delta _{3}=0$ corresponds to the transition
frequency $\protect\omega _{33}$. The three curves are for
increasing pump intensity \textit{(i)} $I_{3}=0.1$, \textit{(ii)}
$I_{3}=1.0$, \textit{(iii)} $I_{3}=10.0$. The arrows indicate the
positions of the expected \textquotedblleft
vacuum-Rabi\textquotedblright\ peaks at $\pm g_{43}$, where
$g_{43}/2\protect\pi =16$ MHz. In all cases, the recycling field
$\Omega _{4}$ is on resonance $\Delta _{4}=0$ and has intensity
$I_{4}=3$.} \label{nvsdelta3}
\end{figure}

To investigate this question, we consider the dependence of the
average intracavity photon number $\bar{n}$ on the detuning
$\Delta _{3}$\ of the pump field $\Omega _{3}$, with the result of
this analysis illustrated in Fig. \ref{nvsdelta3}. For weak
excitation $I_{3}\lesssim 1$ (well below the peak in Fig.
\ref{f_1}(a)), the intracavity photon $\bar{n}$\ is maximized
around $\Delta _{3}=\pm g_{43}$ (and not at $\Delta _{3}=0$) in
correspondence to the eigenvalue structure for the
$g4\leftrightarrow e3$ manifold in presence of strong coupling.
The excited state $e3$ is now represented by a superposition of
the nondegenerate states $|\psi _{\pm }\rangle $ whose energies
are split by the coupling energy $\pm \hbar g_{43}$ . However, for
large pump intensities $I_{3}\sim 10$, this splitting is lost as
the Autler-Townes effect associated with the pump field on the $
g3\leftrightarrow e3$ transition grows to exceed $g_{43}$.

\subsection{Optical spectrum of the cavity emission}

\begin{figure}[htb]
\includegraphics[width=8cm]{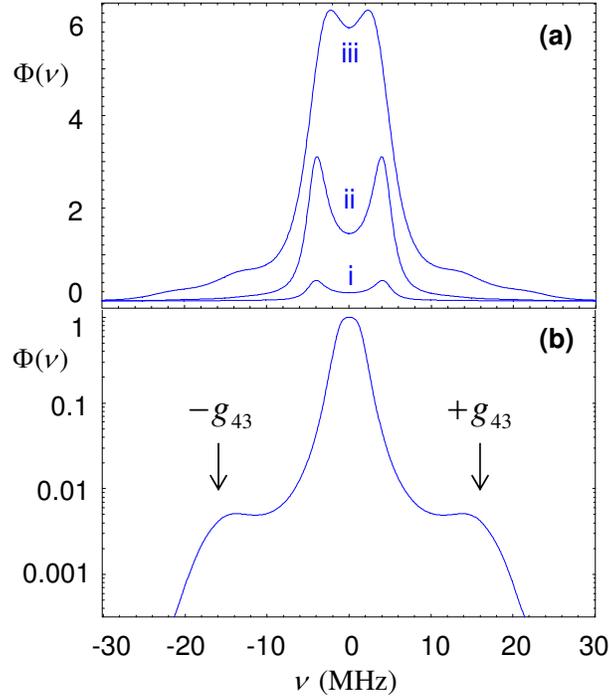}
\caption{The optical spectrum $\Phi (\protect\nu )$ as a function
of frequency offset $\protect\nu $ (in cycles/sec, $\Omega
=2\protect\pi \protect\nu $), where $\protect\nu =0$ corresponds
to the transition frequency $\protect\omega _{43}$. (a) Three
spectra $\Phi (\protect\nu )$ for increasing pump intensity
\textit{(i)} $I_{3}=0.1$, \textit{(ii)} $ I_{3}=1.0$,
\textit{(iii)} $I_{3}=10.0,$ with the recycling intensity $
I_{4}=3$ in all cases. The overall normalization of $\Phi
(\protect\nu )$ is arbitrary, but is common for the three cases.
(b) $\Phi (\protect\nu )$ on a logarithmic scale for decreased
intensities $I_{3}=I_{4}=0.5$, with the peak value of $\Phi$
scaled to unity. The arrows indicate the position of the expected
\textquotedblleft vacuum-Rabi\textquotedblright\ peaks at $\pm
g_{43}$, where $g_{43}/2\protect\pi =16$ MHz. In all cases in (a)
and (b), the pumping field $\Omega _{3}$ and the recycling field
$\Omega _{4}$ are on resonance with their respective transitions
($\Delta _{3}=0=\Delta _{4}$).} \label{oal-spectrum}
\end{figure}

A central feature of a conventional laser is the optical spectrum
of the emitted field, defined by
\begin{equation}
\Phi (\Omega )\equiv \int_{-\infty }^{+\infty }d\tau
\{\lim\limits_{t\rightarrow \infty }\langle \hat{a}^{\dagger
}(t)\hat{a} (t+\tau )\rangle \}\exp (-i\Omega \tau )\text{ ,}
\label{phi}
\end{equation}
where as in Eq. \ref{hamiltonians}, $(\hat{a}^{\dagger },\hat{a})$
are the creation and annihilation operators for the single-mode
field of the cavity coupled to the atomic transition
$e3\leftrightarrow g4$. The results for the Schawlow-Townes
linewidth are well-known and will not be discussed here \cite
{carmichael-book,gardiner-book,sargent-book,haken-book,scully-zubairy}.
Instead, in Fig. \ref{oal-spectrum} we present results specific to
the domain of operation of our system.

For the choice of parameters corresponding to Fig. \ref{f_1}, $\Phi (\Omega
) $ in Fig. \ref{oal-spectrum}(a) exhibits a pronounced two-peak structure,
with the positions of the peaks corresponding to the Autler-Townes splitting
of the ground state by the recycling field $\Omega _{4}$. Contrary to what
might have been expected from the analysis of the previous section, $\Phi
(\Omega )$ shows no distinctive features associated with the vacuum-Rabi
splitting of the excited state. For reduced values of pumping and recycling
intensities $I_{3,4}=0.5$, there are small features in the optical spectrum
at $\Omega \approx \pm g_{43}$, as is illustrated in Fig. \ref{oal-spectrum}
when $\Phi (\Omega )$ is plotted on logarithmic scale. With respect to the
complex degree of coherence \cite{mandel-wolf-95}, the coherence properties
of the light from the one-atom laser in the regime of strong coupling are
set simply by the inverse of the spectral width of $\Phi (\Omega )$, which
can be determined from the plots in Fig. \ref{oal-spectrum}.

The curves shown in Fig. \ref{oal-spectrum} are calculated by way of the
quantum regression theorem applied to the four-state system of Fig.
\ref{setup}.  From the quantum regression theorem, we have that the two
time correlation function in Eq. $\ref{phi}$ is given by
\[
\langle a^\dagger(0) a(\tau)\rangle =
Tr[\rho_{ss}a^\dagger(0)a(\tau)] = Tr[\rho(\tau)a(0)]
\]
where $\rho(\tau)$ is obtained by numerically evolving
\[
\rho_0 = \rho_{ss} a^\dagger(0)
\]
under the master equation, and $\rho_{ss}$ is the steady state
density matrix. By Fourier transforming the correlation function
according to Eq. $\ref{phi}$, we obtain the optical spectrum.

The optical spectrum of the emitted light from our cavity could in
principle be measured by way of heterodyne detection. The cavity
output would be combined on a highly transmissive beam splitter
with a local oscillator beam that is frequency shifted by an
interval $\Delta \omega $ that is large compared to the range of
frequencies in the output field. The optical spectrum is then
obtained by taking the Fourier transform of autocorrelation
function of the resulting heterodyne current. Although we have not
carried out this procedure experimentally, it is straightforward
to model using a quantum jumps simulation of the four state model.
We have computed such spectra for several values of $I_{3}$, using
a local oscillator flux equal to $\kappa $. This is an
experimentally reasonable value, since it is small enough so as to
not saturate the detectors, yet large enough that, as our further
simulations indicate, increasing the flux does not significantly
change the resulting spectrum. The results for the spectrum
obtained from this quantum jumps simulation agree reasonably well
with results from the quantum regression theorem presented in Fig.
\ref{oal-spectrum}.

\section{Quantum theory including Zeeman states and two cavity modes}

In an attempt to provide a more detailed quantitative treatment of our
experiment, we have developed a model that includes all of the Zeeman states
$(F,m_{F})$ for the $F=3,4$ ground levels and the $F^{\prime }=3^{\prime
},4^{\prime }$ excited levels of the $6S_{1/2}\leftrightarrow 6P_{3/2}$
transition in atomic Cesium, of which there are $32$ in total. We also
include two cavity modes with orthogonal linear polarizations to describe
the two nearly degenerate $TEM_{00}$ modes of our cavity \cite{mckeever03},
with three Fock states for each mode $\{|0\rangle ,|1\rangle ,|2\rangle \}$.
The total dimension of the Hilbert space for this set of atomic and field
states is then $d=32\times 3\times 3=288$, making it impractical to obtain
steady state solutions from the master equation directly. Instead, we employ
the \textit{Quantum Optics Toolbox }\cite{tan99} to implement a quantum
jumps simulation, with various expectation values computed from the
stochastic trials.

In broad outline, our expanded model includes Hamiltonian terms of the form
of Eq. \ref{hamiltonians}, with now the terms $\hat{\sigma}_{ij}$
generalized to incorporate each of the various Zeeman states. Likewise, the
coherent coupling of the atom to the cavity takes into account two
orthogonally polarized modes $(\hat{a},\hat{b})$. The operators $\hat{L}_{i}$
are similarly modified to obtain a new master equation that includes the
full set of decay paths among the various states (i.e., $\sigma _{\pm },\pi $
transitions), as well as the associated quantum collapse terms in the
simulation.

We attempt to describe the dynamics arising from the complex state
of spatially varying polarization associated with the $\Omega
_{3,4}$ beams by way of the following simple model. In a
coordinate system with the $x,z$ directions perpendicular to the
cavity axis along $y$, the $\Omega _{3,4}$ beams propagate along
$x,z$ with orthogonal $\sigma _{\pm }$ configurations. The helical
patterns of linear polarization from pairs of counter-propagating
beams then give rise to terms in the interaction Hamiltonians
$\hat{H}_{2,3} $ of the form
\begin{eqnarray}
\hat{H}_{2} &=&\frac{1}{2\sqrt{2}}\Omega
_{3}[(\hat{\Sigma}_{g3,e3}^{z}+\hat{
\Sigma}_{e3,g3}^{z})\sin (\theta _{3x})  \label{H2new} \\
&&+(\hat{\Sigma}_{g3,e3}^{x}+\hat{\Sigma}_{e3,g3}^{x})\sin (\theta _{3z})]
\notag \\
&&+\frac{1}{2}\Omega _{3}[(\hat{\Sigma}_{g3,e3}^{y}+\hat{\Sigma}
_{e3,g3}^{y})(\cos (\theta _{3x})+\cos (\theta _{3z}))]  \notag
\end{eqnarray}
and similarly for $\hat{H}_{3}$ to describe the $\Omega _{4}$
beams with independent phases $(\theta _{4x},\theta _{4z})$. Here
$\Omega_3$ and $ \Omega_4$ are Rabi frequencies corresponding to
the incoherent sum of the intensities of the four individual
beams. In Eq. \ref{H2new}, the operators $
\hat{\Sigma}_{g3,e3}^{x,y,z}$ are linear combinations of various
atomic projection operators for the diverse Zeeman-specific
transitions for linear polarization along $x,y,z$, and are given
explicitly by
\begin{eqnarray}
\hat{\Sigma}^x_{g3,e3} & = & -\frac{1}{\sqrt{2}}(\hat{\Sigma}^{+1}_{g3,e3} -
\hat{\Sigma}^{-1}_{g3,e3}) \\
\hat{\Sigma}^y_{g3,e3} & = & \frac{i}{\sqrt{2}}(\hat{\Sigma}^{+1}_{g3,e3} +
\hat{\Sigma}^{-1}_{g3,e3}) \\
\hat{\Sigma}^x_{g3,e3} & = & \hat{\Sigma}^0_{g3,e3}
\end{eqnarray}
where
\begin{eqnarray}
\hat{\Sigma}^q_{g3,e3} = \sum_m \sum_{m^{\prime}} |g3,m\rangle \langle 3,m;
1, q| 4,m^{\prime}\rangle \langle g4,m^{\prime}|
\end{eqnarray}
The phases $\theta _{i}$ arise from the spatial variations of the
polarization state of the $\Omega _{3,4}$ beams, and are given,
for example, by $\theta _{3x}=k_{3x}x$ with $k_{3x}$ as the wave
vector of the pair of $ \Omega _{3}$ beams propagating along $x$.

The $\Omega _{3,4}$ beams tend to optically pump the atom into
dark states, with this pumping counterbalanced by atomic motion
leading to cooling \cite {boiron96} and by any residual magnetic
field. In our case, imperfections in the FORT\ polarization
\cite{hood01,mckeever03} result in a small pseudo-magnetic field
along the cavity axis $y$ \cite{corwin99} with peak magnitude
$B_{y}^{F}\simeq 0.75$ G. This pseudo-field $B_{y}^{F}$ is
included in our simulations and tends to counteract optical
pumping by the $\Omega _{3,4}$ beams into dark states for linear
polarization in the $x-z$ plane, $\theta_{3x} = \theta_{3z} =
\theta_{4x} = \theta_{4z}=\pi /2$, but has no effect for
polarization along the cavity axis $y$,
$\theta_{3x}=\theta_{3z}=\theta_{4x}=\theta_{4z}=0$.

Overall, the operation of our driven atom-cavity system involves
an interplay of cycling through the levels $g3\rightarrow
e3\rightarrow g4\rightarrow e4\rightarrow g3$ to achieve output
light on the $ e3\rightarrow g4$ transition, and of polarization
gradient cooling for extended trapping times. This latter process
involves atomic motion through the polarization gradients of the
$\Omega _{3,4}$ beams and is greatly complicated by the presence
of $B_{y}^{F}$. The detunings and intensities of the $\Omega
_{3,4}$ beams are chosen operationally such as to optimize the
output from our one-atom laser in a regime of strong coupling,
while at the same time maintaining acceptable trapping times, as
shown in Fig. 2 of Ref. \cite{mckeever03b}.

\subsection{Mean intracavity photon number as a function of pump intensity}

In this section, we present simulation results for the mean intracavity
photon number versus pump intensity. In qualitative terms, we should expect
that the output flux $\kappa \bar{n}$ predicted from the full multi-state
model is significantly below that calculated from the four-state model
presented in Section IV. This is because the atom necessarily spends
increased time in manifolds of dark states associated with the pumping by
the $\Omega _{3,4}$ beams.

\begin{figure*}[htb]
\includegraphics[width=16cm]{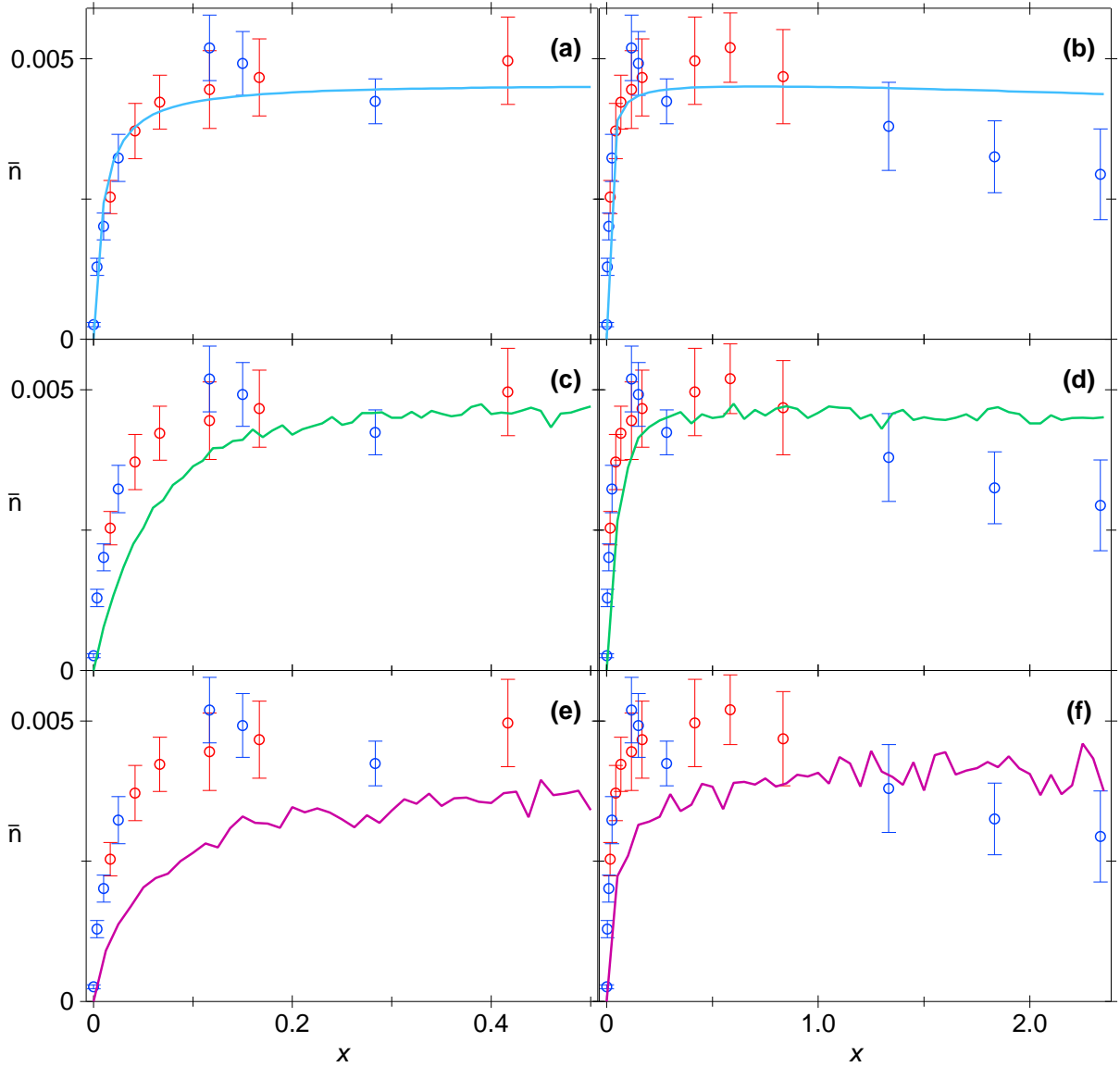}
\caption{Comparison of theory and experiment for the intracavity
photon number $\bar{n}$ as a function of pump intensity $x\equiv
(7/9)(I_3/I_4)$ for fixed $I_{4}=13$ (corresponding to a measured
intensity of $50 mW/cm^2$ ). The measurements (points with error
bars) are from Figure 3 of Ref. \protect\cite{mckeever03b}. (a,b)
$\bar{n}$ versus pump strength $x$ for the four level model with
$\protect\gamma_{43}^{\prime}= 0.07 \times \protect \gamma_{43}$.
(c,d) $\bar{n}$ versus pump strength $x$ for the constant phase
model with $\protect\theta = \protect\pi/2$. (e,f) $\bar{n}$
versus pump strength $x$ for the constant velocity model described
in the text. The immediate onset of emission supports the
conclusion of \textquotedblleft thresholdless\textquotedblright\
lasing. Two independent sets of measurements are shown as the red
and blue points, and agree reasonably well with each other.
Details of the measurements can be found in Ref. \protect\cite
{mckeever03b}, while the parameters for the simulation are given
in the text. } \label{nvsx}
\end{figure*}

We can modify the four level model to account for these effects by
reducing the decay rate $\gamma_{34} \rightarrow
\gamma_{34}^{\prime}$. The slower cycling of the atom due to the
reduction of $\gamma_{34}^{\prime}$ approximates, in a
phenomenological way, the slowing effect on the recycling of the
atom due to optical pumping into dark states. We find that a value
$ \gamma_{34}^{\prime}= 0.07 \times \gamma_{34}$ gives a good fit
to the data (Fig. \ref{nvsx}(a,b)). We plot the intracavity photon
number versus $ x\equiv(7/9)(I_3/I_4)$, since we estimate that
either measured intensity alone is uncertain by a factor of about
$2$, but the ratio is known much more accurately.

For the multi-level simulation, we use two different models to
generate mean intracavity photon number versus pump intensity
curves. In the first model, we neglect the motion of the atom and
attempt to capture the essential features of the optical pumping
processes via a single constant phase $ \theta= \theta _{3x} =
\theta_{3z} = \theta _{4x} =\theta _{4z}$. The choice $\theta =0$
gives no output light, since the $\Omega _{3,4}$ beams pump the
atom into dark states. The value $\theta = \pi/2$ chosen for the
comparison in Fig. \ref{nvsx}(c,d) gives good correspondence
between the simulations and our measurements with the adjustment
of no other parameters. For this curve, we plot the average
$(\bar{n}_a + \bar{n}_b)/2$ of the intracavity field for the two
cavity modes $a$ and $b$.

As a second, more sophisticated model, we assume that the atom moves at a
constant velocity in the radial direction. This gives time dependent phases;
for example, if we assume that the $x$ coordinate of the atom is
\begin{equation*}
x(t) = x_0 + v_x t
\end{equation*}
then
\begin{equation*}
\theta_{3x}(t) = k_{3x} x = \theta_{3x,0} + \omega_{3x} t
\end{equation*}
where $\theta_{3x,0}= k_{3x} x_0$, $\omega_{3x} = k_{3x} v_x$. For
a single simulation run we randomly choose the velocity of the
atom and initial phases of the $\Omega_{3,4}$ pumping beams; the
intensities from 20 such runs are averaged for each value of $x$.
The velocities are chosen uniformly in the range $10 - 20 cm/s$,
which gives angular frequencies in the range $2 \pi 100 - 200
kHz$. The resulting input/output curve is plotted in Fig \ref
{nvsx}(e,f). As before, we plot the average of the intracavity
field for the two cavity modes.

We make no claim for detailed quantitative agreement between
theory and experiment, as the simulations are sensitive to the
parameters which are known only approximately, such as the
intensity of the $\Omega_{3,4}$ pumping beams and the magnitude of
the pseudo and real magnetic fields. Also, the simulations neglect
a number of features of the real system, such as atomic motion in
the axial direction, the dependence of the cavity coupling $g$ on
the position of the atom, and a possible intensity imbalance in
the $\Omega_{3,4}$ pumping beams. However, the simulations do
support the conclusion that the range of coupling values $g$ that
contribute to our results is restricted roughly to
$0.5g_{0}\lesssim g\lesssim g_{0}$. Furthermore, the simulations
yield information about the atomic populations, from which we
deduce that the rate of emission from the cavity $\kappa\bar{n} $
exceeds that by way of fluorescent decay $3^{\prime }\rightarrow
4$, $ \gamma _{4 3^{\prime }}\langle \sigma _{3^{\prime }3^{\prime
}}\rangle $, by roughly tenfold over the range of pump intensity
$I_{3}$ shown in Fig. \ref {nvsx}a.

\subsection{Photon statistics as expressed by the intensity correlation
function $g^{(2)}(\protect\tau )$}

In addition to measurements of $\bar{n}$ versus pumping rate, we
have also investigated the photon statistics of the light emitted
by the TEM$_{00}$ mode of the cavity by way of the two
single-photon detectors \textit{D}$ _{1,2}$ illustrated in Fig. 1
of Ref. \cite{mckeever03b}. From the cross-correlation of the
resulting binned photon arrival times and the mean counting rates
of the signals and the background, we construct the normalized
intensity correlation function (see the \textit{Supplementary
Information} accompanying Ref. \cite{mckeever03b})
\begin{equation}
g^{(2)}(\tau )=\frac{\langle :\hat{I}(t)\hat{I}(t+\tau ):\rangle
}{\langle : \hat{I}(t):\rangle ^{2}}\text{ ,}  \label{g2}
\end{equation}
where the colons denote normal and time ordering for the intensity operators
$\hat{I}$ \cite{mandel-wolf-95}.

\begin{figure*}[htb]
\includegraphics[width=16cm]{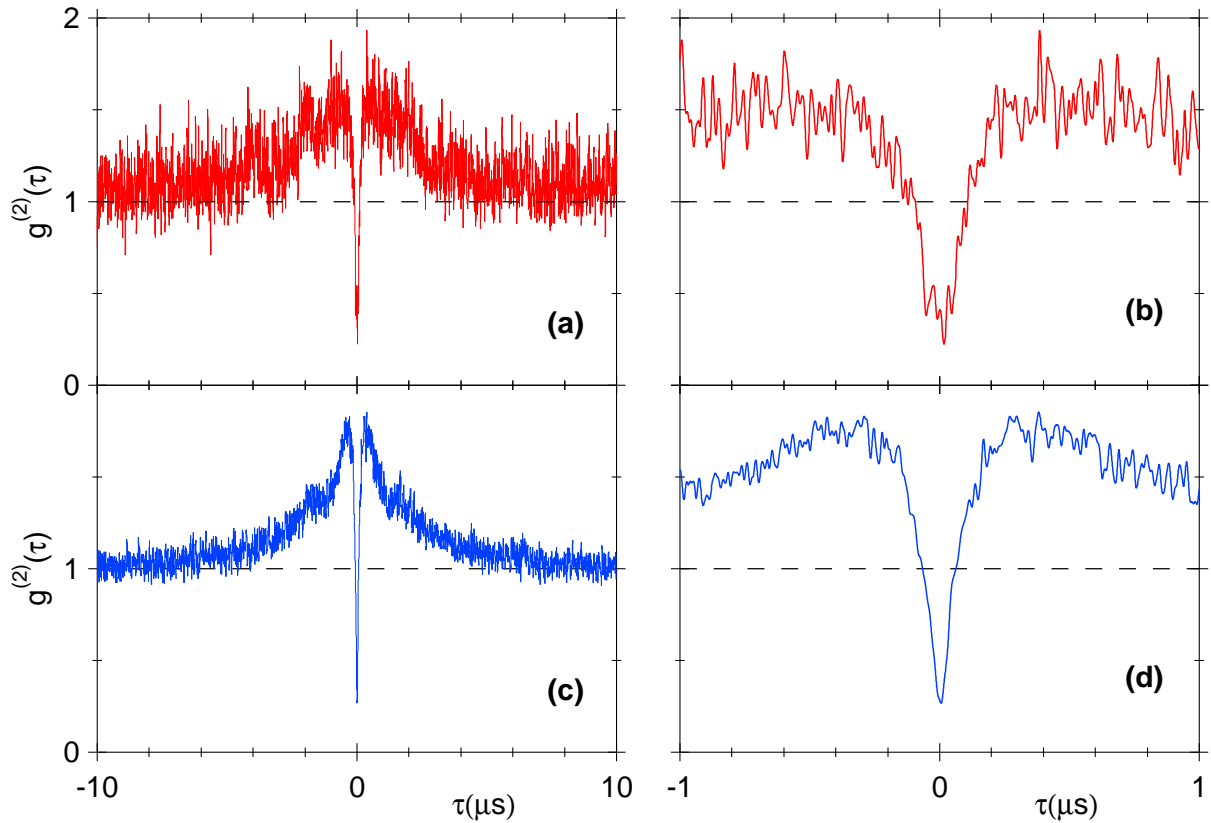}
\caption{The intensity correlation function
$g^{(2)}(\protect\tau)$ of the one-atom laser. (a,b)
$g^{(2)}(\protect\tau)$ for $x \simeq 0.17$ as experimentally
determined in Ref. \protect\cite{mckeever03b}. (c,d) Theoretical
result for $g^{(2)}(\protect\tau)$ for $x = 0.17$ from a quantum
jumps simulation with $\protect\theta=\protect\pi/2$. All traces
have been ``smoothed" by convolution with a Gaussian function of
width $\protect\sigma =5~\mathrm{ns}$.} \label{g2tau-x1}
\end{figure*}

\begin{figure*}[htb]
\includegraphics[width=16cm]{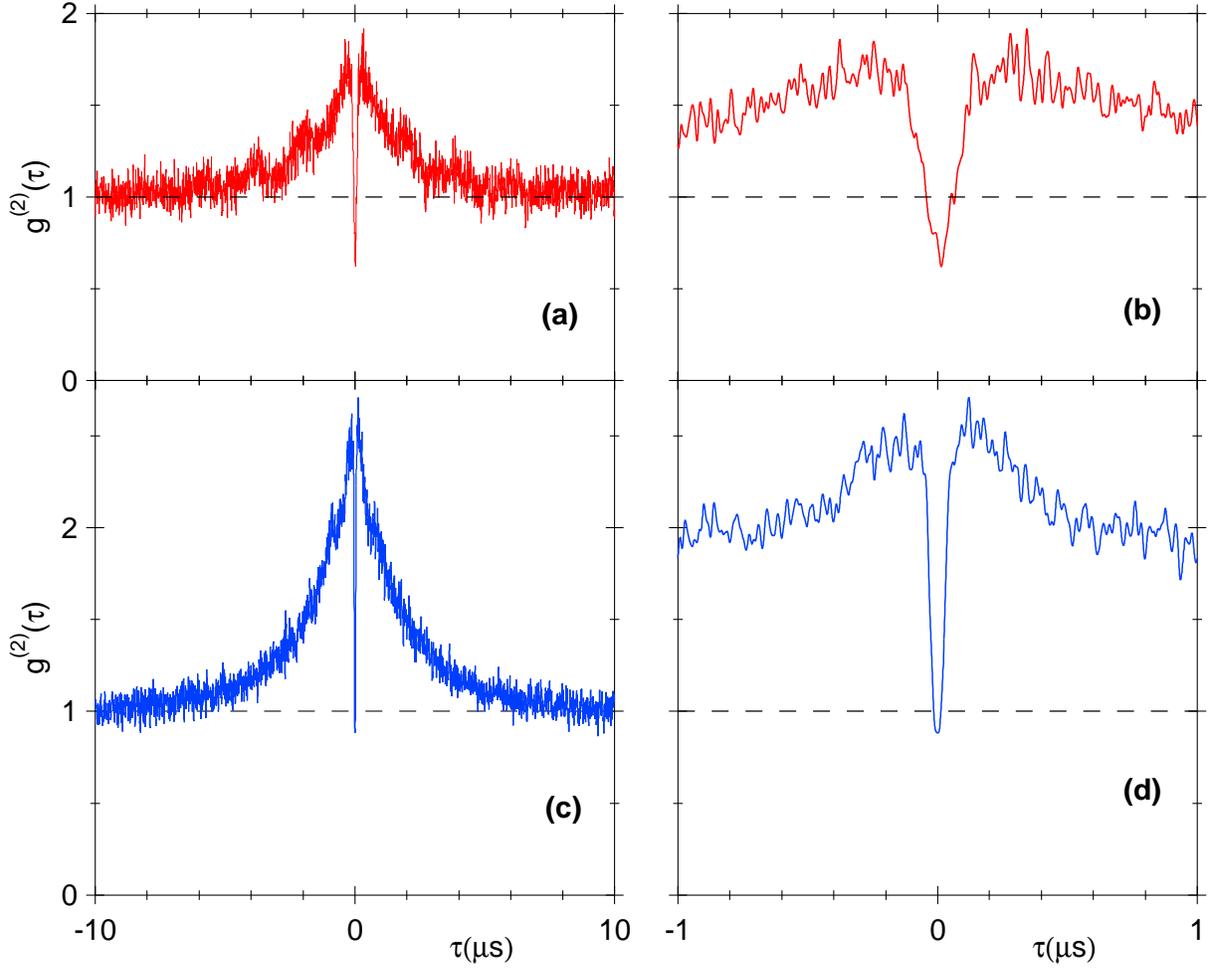}
\caption{The intensity correlation function $g^{(2)}(\protect\tau
)$ of the one-atom laser. (a,b) $g^{(2)}(\protect\tau )$ for
$x\simeq 0.83$ as experimentally determined in Ref.
\protect\cite{mckeever03b}. (c,d) Theoretical result for
$g^{(2)}(\protect\tau )$ for $x=0.83$ from a quantum jumps
simulation with $\protect\theta =\protect\pi /2$. All traces have
been \textquotedblleft smoothed" by convolution with a Gaussian
function of width $\protect\sigma =5~\mathrm{ns}$.}
\label{g2tau-x2}
\end{figure*}

Two measurements for $g^{(2)}(\tau )$ from Figure 4 of Ref. \cite
{mckeever03b} are reproduced in (a,b) of Figs. \ref{g2tau-x1} and
\ref {g2tau-x2}, together with results from our quantum jumps
simulation from the constant phase model with $\theta = \pi/2$, in
(c,d). In Fig. \ref{g2tau-x1}, we again have $I_4\simeq 13$ and
the pump intensity $I_{3}$ is set for operation with $x\simeq
0.17$ near the \textquotedblleft knee\textquotedblright\ in
$\bar{n} $ versus $x$, while in Fig. \ref{g2tau-x2}, the pump
level is increased to $ x\simeq 0.83$. These measurements
demonstrate that the light from the atom-cavity system is
manifestly quantum (i.e., nonclassical) and exhibits photon
antibunching $g^{(2)}(0)<g^{(2)}(\tau )$\ and sub-Poissonian
photon statistics $g^{(2)}(0)<1$ \cite{mandel-wolf-95}. In
agreement with the trend predicted by the four-state model in Fig.
\ref{f_1}(c) (as well as by the full quantum jumps simulation),
$g^{(2)}(0)$ increases with increasing pump intensity, with a
concomitant decrease in these nonclassical effects. The bottleneck
associated with the recycling process leads to this nonclassical
character, since detection of a second photon given the first
detection event requires that the atom be recycled from the $F=4$
ground state back to the $F=3$ ground. In this regard, we point to
the prior work on pump-noise suppressed lasers in multi-level
atomic systems, as for example, in Ref. \cite{ritsch91}.

In more quantitative terms, theoretical results for $g^{(2)}(\tau
)$ from the full quantum jumps simulation are given in parts (c,d)
of Figs. \ref {g2tau-x1} and \ref{g2tau-x2} for $x=0.17$ and
$x=0.83$. The excess fluctuations $g^{(2)}(\tau )\gtrsim 1$
extending over $\tau \simeq \pm 1~ \mathrm{\mu s}$ appear to be
related to the interplay of atomic motion and optical pumping into
dark states \cite{boiron96}, as well as Larmor precession that
arises from residual ellipticity in polarization of the
intracavity FORT \cite{mckeever03,corwin99}.

These results for $g^{(2)}(\tau )$ provide a perspective on the
issue of whether the cavity is effectively \textquotedblleft
empty\textquotedblright\ since $\bar{n}$ is quite small. Based
upon the mean photon flux from the cavity, this is a reasonably
inference, but it is also misleading. The nonzero values for
$g^{(2)}(\tau =0)\simeq 0.3,0.6\gg 0.01$ in Figs. \ref {g2tau-x1}
and \ref{g2tau-x2} are in fact due to the presence of more than
one photon in the cavity. Although the mean intracavity photon
number is only $\bar{n}\sim 0.005$, this number is comparable to
the saturation photon number $n_{0}\simeq 0.013$. Indeed, the
quantum statistical character of the intracavity field is
determined from the self-consistent interplay of atom and cavity
field as in standard laser theories, even though it might appear
as this interplay is not relevant to the determination of a
dynamic steady state. Figure \ref{variousvsf} attempts to
illustrate this point by investigating the passage from the domain
of conventional laser theory through the regime of strong coupling
and thence into a domain of single photon generation with
$g^{(2)}(\tau =0)\simeq 0$ over the entire range of pumping
conditions.

\subsection{Discussion of possible coherence effects}

In Section IIIA we briefly described our analysis of an equivalent
Raman scheme to address the question of possible coherence effects
associated with the $\Omega _{4}$ recycling beam. Beyond this
analysis, we have also considered the possibility that various
other coherent processes associated with the pump fields might be
important. One concern relates to the possibility that $4$-wave
mixing processes could be important, as for example, in a
wave-mixing process that cycles the atom $3\rightarrow 3^{\prime
}\rightarrow 4\rightarrow 4^{\prime }\rightarrow 3$ \cite
{gauthier03}. From an operational perspective, if there were to be
a correlated process involved in the cycling of the atom
$3\rightarrow 3^{\prime }\rightarrow 4\rightarrow 4^{\prime
}\rightarrow 3$, then two photons would be emitted into the cavity
mode (the \textquotedblleft signal\textquotedblright\ on the
$3^{\prime }\rightarrow 4$ transition and the \textquotedblleft
idler\textquotedblright\ on the $4^{\prime }\rightarrow 3$
transition). In this case since we employ no filter to block the
\textquotedblleft idler\textquotedblright\ field separated by
$9.2$GHz, the measured intensity correlation function
$g^{(2)}(\tau )$ for the emitted light from the cavity would
exhibit bunching around $\tau =0$, instead of the observed
antibunching and sub-Poissonian character. The measured character
of $g^{(2)}(\tau )$ therefore argues against a coherent process
that cycles the atom from an initial quantum state and back to
that state by way of coherent processes involving coupling to the
cavity field.

We also note that the coherent coupling of the cavity field and
atom for the $4^{\prime }\rightarrow 3$ transition is greatly
suppressed due to the large detuning $\simeq 9.2$GHz, leading to
an effective coupling coefficient $ (g_{eff}/2\pi \sim 30$
kHz$)\ll (\gamma _{\Vert }/2\pi \simeq 5.2$ MHz$)$. Therefore, for
whatever mixing processes, the coupling to the external vacuum
modes characterized by the rate $\gamma _{\Vert }$ should dominate
that due to $g_{eff}$. In this regard, note that we have included
the effect of off-resonant coupling of the $4^{\prime }$ excited
state in our simulations (which is only $\simeq 200$ MHz detuned).
The relevant process is then excitation $4\rightarrow 4^{\prime }$
via the $\Omega _{4}$ pump field, followed by emission into the
cavity mode due to the coherent coupling of the transition
$4^{\prime }\rightarrow 4$. This coupling increases the
intracavity photon number by only about $10\%$, suggesting that
coupling for the $4^{\prime }\rightarrow 3$ transition $9.2$ GHz
away is negligible.

In support of these comments, our detailed numerical simulations
agree sensibly well with the observed behavior of $g^{(2)}(\tau )$
(as in Figures \ref{g2tau-x1} and \ref{g2tau-x2}), and do not
include any \textquotedblleft wave-mixing\textquotedblright\
effects. This statement is likewise valid for the dependence of
photon number versus pump level $\Omega _{3}^{2}$. Furthermore, as
previously discussed, the model calculation for a four-state
system agrees well in its essential characteristics with a
three-state system where the decay of the ground state
$4\rightarrow 3$ is via an ad hoc spontaneous process (as in a
Raman laser) rather than by pumping $ 4\rightarrow 4^{\prime }$
and decay $4^{\prime }\rightarrow 3$.

A final general comment relates to the nature of phase-matching
(e.g., as applied to $4$-wave mixing and parametric down
conversion) for a single atom in a cavity. For a sample of atoms
(or a crystal), there is a geometry that defines directions for
which fields from successive atoms might add constructively for
various waves (e.g., pump, signal, idler). Cavities can then be
placed around these directions to enhance the processes (e.g., the
threshold for an optical parametric oscillator is reduced by a
factor of the square of the cavity finesse for resonant
enhancement of both signal and idler fields). Clearly a cavity
would be ineffective if its geometry did not match the preferred
geometry defined by the sample and pump beams. However, for a
single atom as in our experiment, these considerations do not
apply in nearly the same fashion. The relevant issues are the
coherent coupling coefficients $g_{ij}$ of the various atomic
transitions to the cavity field.

\section{Summary}

We have presented a simplified four-level model which describes
the qualitative features of our experiment. We have shown how
decreasing the cavity length causes the model system to move from
a regime of weak coupling, where the semiclassical laser theory
applies, into a regime of strong coupling, where quantum
deviations become important. The four-state model predicts many of
the observed features of our experimental system, including the
qualitative shape of the intracavity photon number versus pumping
intensity curve, and photon antibunching.

In addition, to predict quantitative values for comparison with our
experimental results, we have developed a full multi-level model which
correctly describes optical pumping and Larmor precession effects within the
Zeeman substructure. We have shown that these effects play an important role
in describing the observed input/ output characteristics of the system, and
that by including a simple model for the motion of the atom we can obtain
reasonable agreement with the experimentally observed curve. We have also
used the simulation to calculate intensity correlation functions, and have
compared these results to measurements of $g^{(2)}(\tau )$ from our
experiment.

We gratefully acknowledge interactions with K. Birnbaum, L.-M. Duan, D. J.
Gauthier, T. Lynn, T. Northup, A. S. Parkins, and D. M. Stamper-Kurn. This
work was supported by the National Science Foundation, by the Caltech MURI
Center for Quantum Networks under ARO Grant No. DAAD19-00-1-0374, and by the
Office of Naval Research.


\begin{thebibliography}{99}
\bibitem{mckeever03b} J.~McKeever, A.~Boca, A.~D. Boozer, J.~R. Buck, and
H.~J. Kimble, Nature (London) \textbf{425}, 268 (2003).

\bibitem{mu92} Y.~Mu and C.~M. Savage, Phys. Rev. A \textbf{46}, 5944 (1992).

\bibitem{ginzel93} C.~Ginzel, H.-J. Briegel, U.~Martini, B.-G. Englert, and
A.~Schenzle, Phys. Rev. A \textbf{48}, 732 (1993).

\bibitem{pellizzari94a} T.~Pellizzari and H.~Ritsch, Phys. Rev. Lett.
\textbf{72}, 3973 (1994).

\bibitem{pellizzari94b} T.~Pellizzari and H.~Ritsch, J. Mod. Opt. \textbf{41}
, 609 (1994).

\bibitem{horak95} P.~Horak, K.~M. Gheri, and H.~Ritsch, Phys. Rev. A \textbf{
51}, 3257 (1995).

\bibitem{briegel96} H.-J. Briegel, G.~M. Meyer, and B.-G. Englert, Phys.
Rev. A \textbf{53}, 1143 (1996).

\bibitem{meyer97a} G.~M. Meyer, H.-J. Briegel, and H.~Walther, Europhys.
Lett. \textbf{37}, 317 (1997).

\bibitem{loeffler97} M.~L\"{o}ffler, G.~M. Meyer, and H.~Walther, Phys. Rev.
A \textbf{55}, 3923 (1997).

\bibitem{meyer97b} G.~M. Meyer, M.~L\"{o}ffler, and H.~Walther, Phys. Rev. A
\textbf{56}, R1099 (1997).

\bibitem{meyer98} G.~M. Meyer and H.-J. Briegel, Phys. Rev. A \textbf{58},
3210 (1998).

\bibitem{jones99} B.~Jones, S.~Ghose, J.~P. Clemens, P.~R. Rice, and
L.~M.~Pedrotti, Phys. Rev. A \textbf{60}, 3267 (1999).

\bibitem{chough00} Y.-T. Chough, H.-J. Moon, H.~Nha, and K.~An, Phys. Rev. A
\textbf{63}, 013804 (1996).

\bibitem{fidio01} C.~Di~Fidio, W.~Vogel, R.~L. de Matos Filho, and
L.~Davidovich, Phys. Rev. A \textbf{65}, 013811 (2001).

\bibitem{kilin02} S.~Ya. Kilin and T.~B. Karlovich, JETP \textbf{95}, 805
(2002).

\bibitem{rice03} J. P. Clemens, P. R. Rice, and L. M. Pedrotti (2003).

\bibitem{salzburger03} T. Salzburger and H. Ritsch, quant-ph/0312181.

\bibitem{demartini88} F.~De~Martini and G.~R. Jacobivitz, Phys. Rev. Lett.
\textbf{60}, 1711 (1988).

\bibitem{rice94} P.~R. Rice and H.~J. Carmichael, Phys. Rev. A \textbf{50},
4318 (1994).

\bibitem{jin94} R.~Jin, D.~Boggavarapu, M.~Sargent III, P.~Meystre, H.~M.
Gibbs, and G.~Khitrova, Phys. Rev. A \textbf{49}, 4038 (1994).

\bibitem{bjork94} G.~Bj\"{o}rk, A.~Karlsson, and Y.~Yamamoto, Phys. Rev. A
\textbf{50}, 1675 (1994).

\bibitem{protsenko99} I.~Protsenko, P.~Domokos, V.~Lefevre-Seguin, J.~Hare,
J.~M. Raimond, and L.~Davidovich, Phys. Rev. A \textbf{59}, 1667 (2001).

\bibitem{eberly83} J. J. Sanchez-Mondragon, N. B. Narozhny, and J. H.
Eberly, Phys. Rev. Lett. \textbf{51}, 550 (1983).

\bibitem{agarwal84} G. S. Agarwal, Phys. Rev. Lett. \textbf{53}, 1732 (1984).

\bibitem{thompson92} R. J. Thompson, G. Rempe, and H. J. Kimble, Phys. Rev.
Lett. \textbf{68}, 1132 (1992).

\bibitem{berman94} \textit{Cavity Quantum Electrodynamics}, edited by
P.~Berman (Academic Press, San Diego, 1994).

\bibitem{meystre92} \textit{Cavity Quantum Optics and the Quantum Measurement
}, P.~Meystre, in \textit{Progress in Optics, Vol. XXX} edited by E.~Wolf
(Elsevier Science Publishers B.V., Amsterdam, 1992), pp. 261-355.

\bibitem{yamamoto-slusher93} Y.~Yamamoto and R.~E. Slusher, Phys. Today
\textbf{46}(6), 66 (1993).

\bibitem{chang-campillo96} \textit{Optical Processes in Microcavities},
edited by R.~K. Chang and A.~J. Campillo, (World Scientific, Singapore,
1996).

\bibitem{vahala03} K.~J. Vahala, Nature (London) \textbf{424}, 839 (2003).

\bibitem{carmichael-book} H.~J. Carmichael, \textit{Statistical Methods in
Quantum Optics 1} (Springer-Verlag, Berlin, 1999).

\bibitem{gardiner-book} C.~W. Gardiner and P.~Zoller, \textit{Quantum Noise}
(Springer-Verlag, Berlin, 2000).

\bibitem{hjk sweden} H.~J. Kimble, Physica Scripta \textbf{T76}, 127 (1998).

\bibitem{sargent-book} M.~Sargent III, M.~O. Scully, and W.~E. Lamb Jr.,
\textit{Laser Physics} (Addison-Wesley, Reading Mass., 1974).

\bibitem{eschmann99} A.~Eschmann and R.~J. Ballagh, Phys. Rev. A \textbf{60}
, 559 (1999).

\bibitem{tan99} S.~M. Tan, J. Opt. B: Quantum Semiclass. Opt. \textbf{1},
424 (1999).

\bibitem{mandel-wolf-95} L.~Mandel and E.~Wolf, \textit{Optical Coherence
and Quantum Optics} (Cambridge University Press, New York, 1995).

\bibitem{haken-book} H.~Haken, \textit{Laser Theory} (Springer Verlag,
Berlin, 1984).

\bibitem{scully-zubairy} M.~O. Scully and M.~S. Zubairy, \textit{Quantum
Optics} (Cambridge University Press, Cambridge, 1997).

\bibitem{lugiato} \textit{Theory of Optical Bistability}, L. A. Lugiato, in
\textit{Progress in Optics, Vol. XXI} edited by E.~Wolf (Elsevier Science
Publishers B.V., Amsterdam, 1984), pp. 69-216.

\bibitem{turchette95} Q.~A. Turchette, R.~J. Thompson, and H.~J. Kimble,
Appl. Phys. B \textbf{60}, S1 (1995).

\bibitem{hood01} C.~J. Hood, H.~J. Kimble, and J.~Ye, Phys. Rev. A \textbf{64
}, 033804 (2001).

\bibitem{mckeever03} J.~McKeever, J.~R. Buck, A.~D. Boozer, A.~Kuzmich,
H.-C. N\"{a}gerl, D.~M. Stamper-Kurn, H.~J. Kimble, Phys. Rev.
Lett. \textbf{ 90}, 133602 (2003).

\bibitem{boiron96} D.~Boiron, A.~Michaud, P.~Lemonde, Y.~Castin, and
C.~Salomon, Phys. Rev. A \textbf{53}, R3734 (1996) and references therein.

\bibitem{corwin99} K.~L. Corwin, S.~J.~M. Kuppens, D.~Cho, and C.~E. Wieman,
Phys. Rev. Lett. \textbf{83}, 1311 (1999).

\bibitem{gauthier03} The discussion about possible coherent wave-mixing
effects was initiated by D. J. Gauthier, to whom we are most grateful.

\bibitem{ritsch91} H. Ritsch, P. Zoller, C. W. Gardiner, and D. F. Walls,
Phys. Rev. A \textbf{44}, 3361 (1991), and references therein.
\end{thebibliography}
\end{document}